\documentclass[iop,preprint2]{emulateapj1}
\usepackage{psfig}
\usepackage{apjfonts}
\usepackage{ulem}
\usepackage[dvips]{color}
\usepackage{graphicx}
\usepackage[USenglish]{babel}
\usepackage{mathrsfs}
\usepackage[pdfa]{hyperref}
\usepackage{color}
\def\kms{{\rm km}\,{\rm s}^{-1}}
\def\masyr{{\rm mas}\,{\rm yr}^{-1}}

\def\mua{\mu_{\alpha}\cos\delta}
\def\mud{\mu_{\delta}}
\def\mut{\mu_{\rm TAN}}

\begin{document}

\title{Hubble Space Telescope Proper Motion (HSTPROMO) Catalogs of
  Galactic Globular Clusters. V.  The rapid rotation of 47~Tuc traced
  and modeled in three dimensions$^{\ast}$}\footnotetext[$^{\ast}$]
      {Based on archival observations with the NASA/ESA \textit{Hubble
          Space Telescope}, obtained at the Space Telescope Science
        Institute, which is operated by AURA, Inc., under NASA
        contract NAS 5-26555.}
\received{February 27, 2017}
\accepted{June 26, 2017}
\author{
A.\ Bellini\altaffilmark{1},
P.\ Bianchini\altaffilmark{2,3},
A.\ L.\ Varri\altaffilmark{4},
J. Anderson\altaffilmark{1},
G.\ Piotto\altaffilmark{5,6},
R.\ P.\ van der Marel\altaffilmark{1}, 
E.\ Vesperini\altaffilmark{7}, and
L.\ L.\ Watkins\altaffilmark{1}
}

\altaffiltext{1}{Space Telescope Science Institute, 3700 San Martin
  Dr., Baltimore, MD 21218, USA}
\altaffiltext{2}{Max-Planck Institute for Astronomy, Koenigstuhl 17,
  D-69117 Heidelberg, Germany}
\altaffiltext{3}{McMaster:\ Department of Physics and Astronomy,
  McMaster University, Hamilton, Ontario, L8S 4M1, Canada}
\altaffiltext{4}{Institute for Astronomy, University of Edinburgh,
  Royal Observatory, Blackford Hill, Edinburgh EH9 3HJ, UK}
\altaffiltext{5}{Dipartimento di Fisica e Astronomia ``Galileo
  Galilei'', Universit\`a di Padova, v.co dell'Osservatorio 3,
  I-35122, Padova, Italy}
\altaffiltext{6}{Istituto Nazionale di Astrofisica, Osservatorio
  Astronomico di Padova, v.co dell'Osservatorio 5, I-35122, Padova,
  Italy}
\altaffiltext{7}{Department of Astronomy, Indiana University,
  Bloomington, IN 47405, USA}
\topmargin 1.0cm

\begin{abstract} 
High-precision proper motions of the globular cluster 47~Tuc have
allowed us to measure for the first time the cluster rotation in the
plane of the sky and the velocity anisotropy profile from the cluster
core out to about 13$^\prime$. These profiles are coupled with prior
measurements along the line of sight and the surface-brightness
profile, and fit all together with self-consistent models specifically
constructed to describe quasi-relaxed stellar systems with realistic
differential rotation, axisymmetry and pressure anisotropy. The
best-fit model provides an inclination angle $i$ between the rotation
axis and the line-of-sight direction of $30^\circ$, and is able to
simultaneously reproduce the full three-dimensional kinematics and
structure of the cluster, while preserving a good agreement with the
projected morphology. Literature models based solely on line-of-sight
measurements imply a significantly different inclination angle
($i=45^\circ$), demonstrating that proper motions play a key role in
constraining the intrinsic structure of 47~Tuc. Our best-fit global
dynamical model implies an internal rotation higher than previous
studies have shown, and suggests a peak of the intrinsic $V/\sigma$
ratio of $\sim$0.9 at around two half-light radii, with a
non-monotonic intrinsic ellipticity profile reaching values up to
0.45. Our study unveils a new degree of dynamical complexity in
47~Tuc, which may be leveraged to provide new insights into the
formation and evolution of globular clusters.
\end{abstract}


\maketitle

\section{Introduction}
\label{sec:intro}

Globular-cluster (GC) formation, internal dynamical evolution and the
effects of the external tidal field of the host galaxy are expected to
leave a number of fingerprints on a cluster's structural, 
morphological and kinematical properties.

Until recently, the dynamical characterization of most GCs has been
limited to the surface-brightness or projected-star-count radial
profiles. These profiles can shed light on some aspects of cluster
dynamical evolution -- such as the identification of the systems that
have already evolved past the core collapse phase (see, e.g.,
\citealt{1986ApJ...305L..61D, 1989ApJ...339..904C,
  1994AJ....108.1292D, Trager1995}), and to calculate a number of
fundamental global properties (see, e.g.,
\citealt{2005ApJS..161..304M, 2013ApJ...774..151M}) -- but they
provide only a very partial view of a cluster's dynamical state. To
construct a complete dynamical picture of clusters, we require an
accurate understanding of both their internal kinematics and their
detailed morphologies. Such information places key constraints on
theoretical studies, and thus allows us to reconstruct cluster
dynamical histories, and determine the role played by different
dynamical processes \citep[e.g.,][]{2011MNRAS.410.2698G}.

\begin{figure*}[ht!]
\centering
\includegraphics[angle=90,width=\textwidth]{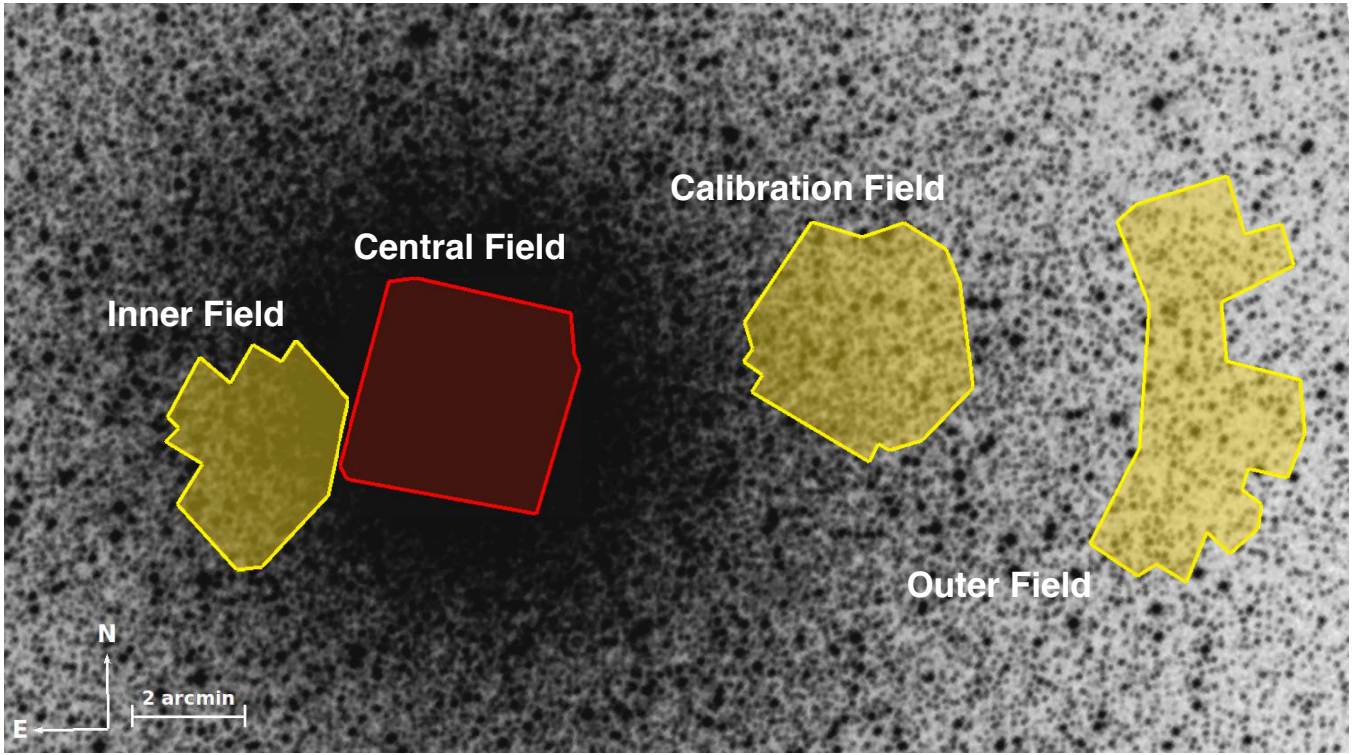}\\
\caption{The footprints of the four fields analyzed in this work,
  superimposed on an SDSS image of 47~Tuc. The irregular shapes of the
  fields are due to overlapping datasets with different pointings,
  rotation angles, and sometimes even taken with different
  detectors. The proper-motion catalog of the central field (in red)
  is that of Paper~I, while the other three fields in yellow (inner,
  calibration and outer, from left to right, respectively) have been
  specifically reduced and analyzed to measure the rotation of 47~Tuc
  in the plane of the sky. The scale and orientation are also shown on
  the bottom-left corner of the figure.\\~\\}
\label{f:fov}
\end{figure*}

After some early pioneering work on cluster kinematical properties
(see, e.g., \citealt{1993ASPC...50..357P, 2000A&A...360..472V,
  1997A&ARv...8....1M} and references therein), there has been a
recent revival in observational studies of internal cluster kinematics
based on ESO/VLT radial velocities (see, e.g.,
\citealt{2012A&A...538A..18B, 2013ApJ...769..107L,
  2016A&A...588A.149K}) and \textit{Hubble Space Telescope}
(\textit{HST})-based, proper-motion (PM) measurements (see, e.g.,
\citealt{bel14pm, 2015ApJ...803...29W}, hereafter Paper~I and
Paper~II, respectively).

These kinematics have already been used to address a number of
fundamental issues, including the existence of intermediate-mass black
holes at the centers of GCs (see, e.g., \citealt{2010ApJ...719L..60N,
  2010ApJ...710.1032A, 2010ApJ...710.1063V, 2013ApJ...769..107L,
  2013A&A...552A..49L}), the kinematical differences between multiple
stellar populations (\citealt{2013ApJ...771L..15R,
  2015ApJ...810L..13B, 2017MNRAS.465.3515C}), differences between the
velocity dispersion of stars with different masses and the degree of
energy equipartition (see, e.g., \citealt{2016ApJ...827...12B},
Paper~IV), the presence and origin of anisotropy in the velocity
distribution (Paper~II), and the strength of cluster rotation and its
possible link with the cluster morphology (see, e.g.,
\citealt{2012A&A...538A..18B, paolo13, Kacharov2014,
  2014ApJ...787L..26F, 2015A&A...573A.115L, bob17}). Forthcoming data
from the \textit{Gaia} mission (see, e.g.,
\citealt{2017MNRAS.467..412P}) will further enrich the observational
landscape.

On the theoretical side, renewed efforts are being made to expand the
numerical and analytical tools necessary to interpret the results of
these observational studies and gather a deeper understanding of their
implications in the context of GC dynamical evolution.  These efforts
include studies proposing new distribution-function-based models of
rotating and anisotropic models (see, e.g., \citealt{VarriBertin2012,
  2015MNRAS.454..576G, 2016AA...590A..16D}) extending the widely used
(but limited to spherical symmetry and isotropic velocity
distribution) models such as \citet{1966AJ.....71...64K} or
\citet{1975AJ.....80..175W} models, and numerous numerical studies
exploring, for example, the evolution of rotating and anisotropic
models (see, e.g., the early studies of \citealt{1999MNRAS.302...81E,
  2002MNRAS.334..310K, 2007MNRAS.377..465E}, and the more recent
investigations by \citealt{2013MNRAS.430.2960H, 2016MNRAS.461..402T,
  t2017, 2016MNRAS.462..696Z}), the evolution of the degree of energy
equipartition (see, e.g., \citealt{TrentivanderMarel2013,
  Bianchini2016, 2017MNRAS.464.1977W}) the role of potential escapers
in the observed velocity dispersion profiles (see, e.g.,
\citealt{2010MNRAS.407.2241K, 2017MNRAS.466.3937C}) and the
kinematical implications of star loss from GCs (see, e.g.,
\citealt{1975AJ.....80..290K, 2016MNRAS.455.3693T}).

On the observational, proper-motion-based side, measurements of GC
rotation in the plane of the sky had been completely lacking until
\citet{2000A&A...360..472V} took advantage of a large number of
photographic plates spanning half a century to measure a
plane-of-the-sky differential rotation for the GC $\omega$~Centauri
(see their Fig.~18). A few years later, \citet{ak47t} took advantage
of the high-precision astrometric capabilities of \textit{HST} to
obtain a direct measurement of the plane-of-the-sky rotation of 47~Tuc
(NGC~104) using two diametrically opposite fields $\sim$5$^\prime$
from the cluster center.  The authors exploited the unique feature of
47~Tuc to have a large number of relatively bright Small Magellanic
Cloud (SMC) stars in the background, and SMC stars themselves could be
used as reference objects. In contrast, much fainter galaxies or much
rarer QSOs must be used as reference objects for the vast majority of
the other GCs. A few other GCs analogous to 47~Tuc are:\ (1) NGC~362,
which is also in front of SMC stars and will be the subject of a
forthcoming paper in this series, (2) NGC~6652, which is in front of
Sagittarius Dwarf Spheroidal (Sgr dSph) galaxy stars, and (3)
NGC~6681, also in front of Sgr dSph stars. The plane-of-the-sky
rotation of NGC~6681 has indeed been recently measured (with a null
result) by \citet{2013ApJ...779...81M}, but using background galaxies
as a reference. The \citet{2013ApJ...779...81M} work represents the
third --and so far the last-- observational, PM-based work on the
subject.

In this paper, we extend the pioneering work of \citet{ak47t} and
measure, for the first time, the rotation curve of 47~Tuc in the plane
of the sky from the center of the cluster out to about 13$^\prime$
(about 4 half-light radii, \citealt{h96}). We present
state-of-the-art, high-precision, \textit{HST}-based PM measurements
for the cluster, which we combine with existing line-of-sight (LOS)
velocities to determine the strength of the cluster present-day
rotation and anisotropy in the velocity distribution. By combining the
kinematical data with the known surface-brightness profile, we build a
detailed self-consistent model of the cluster. This work represents
the most complete dynamical characterization of any Galactic GC and
clearly illustrates how only such a detailed study can reveal the
intrinsic dynamical properties of a cluster and provides the needed
constraints to explore the possible evolutionary paths leading to the
present-day observed properties.

The structure of this paper is the following:\ in
Sect.~\ref{sec:dataset} we present the \textit{HST} data set and the
reduction techniques used to measure high-precision PMs. The results
concerning the PM-based rotation of 47~Tuc in the plane of the sky and
the velocity anisotropy are presented, respectively, in
Sect.~\ref{sec:rot} and \ref{s:vdap}. In Sect.~\ref{sec:theo} we
present the results of a detailed dynamical model fitting to the
observational data using the distribution-function based models of
\citet{VarriBertin2012}. Conclusions are summarized in
Sect.\ref{sec:concl}.

\section{Data sets and Reduction}
\label{sec:dataset}

We measured PMs in four fields located at different radial distances
from the center of 47~Tuc. These fields are roughly aligned along the
same axis with respect to the cluster's center, and span over 1200
arcsecs ($>$$20^\prime$) on the sky. In Fig.~\ref{f:fov} we show the
footprints of these four fields superimposed on a Sloan Digital Sky
Survey (SDSS) image of the cluster. The four fields are hereafter
identified, from the East to the West, as:\ the inner field, the
central field, the calibration field, and the outer field. We made use
of exposures taken through three different \textit{HST} cameras:\ the
Wide-Field Planetary Camera 2 (WFPC2), both the High-Resolution
Channel (HRC) and the Wide-Field Channel (WFC) of the Advanced Camera
for Surveys (ACS), and the Ultraviolet-VISible (UVIS) channel of the
Wide-Field Camera 3 (WFC3).

The data reduction and analysis are based on \texttt{\_flt}-type
exposures (or the equivalent \texttt{\_c0f} format for WFPC2), as they
preserve the un-resampled pixel data for optimal stellar-profile
fitting. All ACS/WFC and WFC3/UVIS exposures were corrected for
charge-transfer-efficiency (CTE) defects \citep{an10}.

Images were then reduced using the software family \texttt{img2xym}
\citep{an06a}, employing either single (HRC), spatially-varying
(WFPC2), or spatially- and time-dependent empirical PSFs (WFC and
UVIS, \citealt{2000PASP..112.1360A, an06a, an06b,
  2013ApJ...769L..32B}).  Stellar positions were corrected for
geometric distortion using the state-of-the-art solutions provided by
\citet{2003PASP..115..113A, an06a, an06b, bb09, b11}.  The photometric
data of the central field comes directly from
\citet{2008AJ....135.2055A}. Photometry of the other fields was
calibrated following the prescriptions given in
\citet{2005PASP..117.1049S}.

Stellar positions in each exposure were transformed into a common,
distortion-free, reference frame (the master frame). Only stars
measured in at least four distinct exposures with a time baseline of
at least six months were considered for the present analysis. PMs were
computed using the \textit{central overlap} method
\citep{1971PMcCO..16..267E}, in which each exposure is considered as a
stand-alone epoch. In a nutshell, we transformed stellar positions 
as measured on the individual exposures on the master frame, by means
of general, 6-parameter linear transformations. For each star, its
master-frame transformed positions as a function of the epoch are
fitted by a straight line, the slope of which gives us a direct
measurement of the stellar motion. We applied a careful data-rejection
procedure to remove outliers or mismatches (see Section~5.5 of Paper~I
for more details).  Extensive simulations have demonstrated the
reliability of both estimated PMs and PM errors.

In the following, we describe the reduction procedures of each field.

\subsection{The central field}
\label{ss:data:cenfld}

The PM catalog of the central field is that published in Paper~I.  We
analyzed 433 exposures taken with the ACS (both HRC and WFC channels)
and the WFC3/UVIS. The complete list of observations can be found in
Table~7 of Paper~I. The available time baseline used to compute the
motion of each star goes from 0.52 to 10.32 years (median of 8.23),
depending on the available overlap between different exposures.  We
have 103$\,$638 stars with measured PMs in the central field. We
closely applied the prescriptions given in Section~7.5 of Paper~I in
order to select only high-quality PM measurements (53$\,$898 stars).

\subsection{The calibration field}
\label{ss:data:calfld}

This particular field was selected as one of the
instrument-calibration fields for the ACS and WFC3 detectors, and has
been repeatedly observed since 2002, when ACS was installed on-board
\textit{HST}. We chose a subsample of deep exposures taken from 2002
to 2014 with ACS/WFC and WFC3/UVIS (see Table~1).  The final PM
catalog contains 13$\,$466 objects, 12$\,$132 of which passed our
high-quality selection criteria.

\begin{table}[t!]
\centering
\footnotesize{
\begin{tabular}{ccccl}
\multicolumn{5}{c}{\textbf{Table~1}}\\
\multicolumn{5}{c}{\textsc{List of Observations used for the Calibration Field}}\\
\hline\hline
Epoch&GO&Instr.&Filter&$N\times{\rm exp.time}$\\
\hline
2002.30 & 9018  &   ACS/WFC  & F606W & $1\times765\,{\rm s}$, $1\times1200\,{\rm s}$\\
        &       &            & F814W & $1\times690\,{\rm s}$, $1\times1020\,{\rm s}$\\
\hline
2006.67 & 10730 &   ACS/WFC  & F435W & $8\times350\,{\rm s}$\\
        & 10737 &   ACS/WFC  & F435W & $1\times339\,{\rm s}$\\
        &       &            & F555W & $1\times339\,{\rm s}$\\
        &       &            & F606W & $1\times339\,{\rm s}$\\
        &       &            & F814W & $1\times339\,{\rm s}$\\
\hline
2009.54 & 11444 &  WFC3/UVIS & F606W & $24\times350\,{\rm s}$\\
        & 11452 &  WFC3/UVIS & F814W & $6\times350\,{\rm s}$\\
\hline
2012.22 & 12692 &  WFC3/UVIS & F606W & $2\times350\,{\rm s}$\\
\hline
2014.35 & 13596 & ACS/WFC    & F435W & $2\times339\,{\rm s}$\\
        &       &            & F475W & $2\times339\,{\rm s}$\\
        &       &            & F606W & $2\times339\,{\rm s}$\\
        &       &            & F775W & $2\times339\,{\rm s}$\\
        &       &            & F814W & $2\times339\,{\rm s}$\\
        &       &            & F850L & $2\times339\,{\rm s}$\\
\hline\hline
&&&&\\
\end{tabular}}
\end{table}

\begin{table}[t!]
\centering
\footnotesize{
\begin{tabular}{ccccl}
\multicolumn{5}{c}{\textbf{Table~2}}\\
\multicolumn{5}{c}{\textsc{List of Observations used for the Inner Field}}\\
\hline\hline
Epoch&GO&Instr.&Filter&$N\times{\rm exp.time}$\\
\hline
1994.75 & 5370 & WFPC2      & F606W & $4\times600\,{\rm s}$, $7\times1000\,{\rm s}$\\
        &      &            & F814W & $7\times1000\,{\rm s}$\\
\hline
1994.86 & 5369 & WFPC2      & F606W & $9\times300\,{\rm s}$, $1\times500\,{\rm s}$\\
\hline
1999.79 & 8095 & WFPC2      & F606W & $1\times400\,{\rm s}$, $7\times500\,{\rm s}$\\
\hline
2013.34 & 12971 & ACS/WFC   & F435W & $1\times290\,{\rm s}$, $1\times690\,{\rm s}$\\
        &       &           & F555W & $1\times360\,{\rm s}$, $1\times660\,{\rm s}$\\
\hline\hline
&&&&\\
\end{tabular}}
\end{table}

\begin{table}[t!]
\centering
\footnotesize{
\begin{tabular}{ccccl}
\multicolumn{5}{c}{\textbf{Table~3}}\\
\multicolumn{5}{c}{\textsc{List of Observations of the Outer Field}}\\
\hline\hline
Epoch&GO&Instr.&Filter&$N\times{\rm exp.time}$\\
\hline
2002.35 & 9318 & WFPC2      & F606W & $1\times300\,{\rm s}$, $1\times400\,{\rm s}$\\
        &      &            &       & $7\times500\,{\rm s}$\\
\hline
2002.40 & 8059 & WFPC2      & F450W & $1\times300\,{\rm s}$\\
        &      &            & F606W & $1\times160\,{\rm s}$\\
        &      &            & F814W & $1\times100\,{\rm s}$,$1\times300\,{\rm s}$\\
\hline
2003.40 & 9709 & WFPC2      & F606W & $1\times400\,{\rm s}$,$7\times500\,{\rm s}$\\
\hline
2010.34 & 11677& WFC3/UVIS  & F390W & $1\times1048\,{\rm s}$,$1\times1099\,{\rm s}$\\
        &      &            &       & $1\times1212\,{\rm s}$,$1\times1355\,{\rm s}$\\
        &      &            &       & $2\times1400\,{\rm s}$\\
        &      &            & F606W & $1\times1252\,{\rm s}$,$3\times1347\,{\rm s}$\\
        &      &            &       & $1\times1398\,{\rm s}$,$1\times1402\,{\rm s}$\\
\hline\hline
&&&&\\
\end{tabular}}
\end{table}

\begin{figure*}[ht!]
\centering
\includegraphics[width=\textwidth]{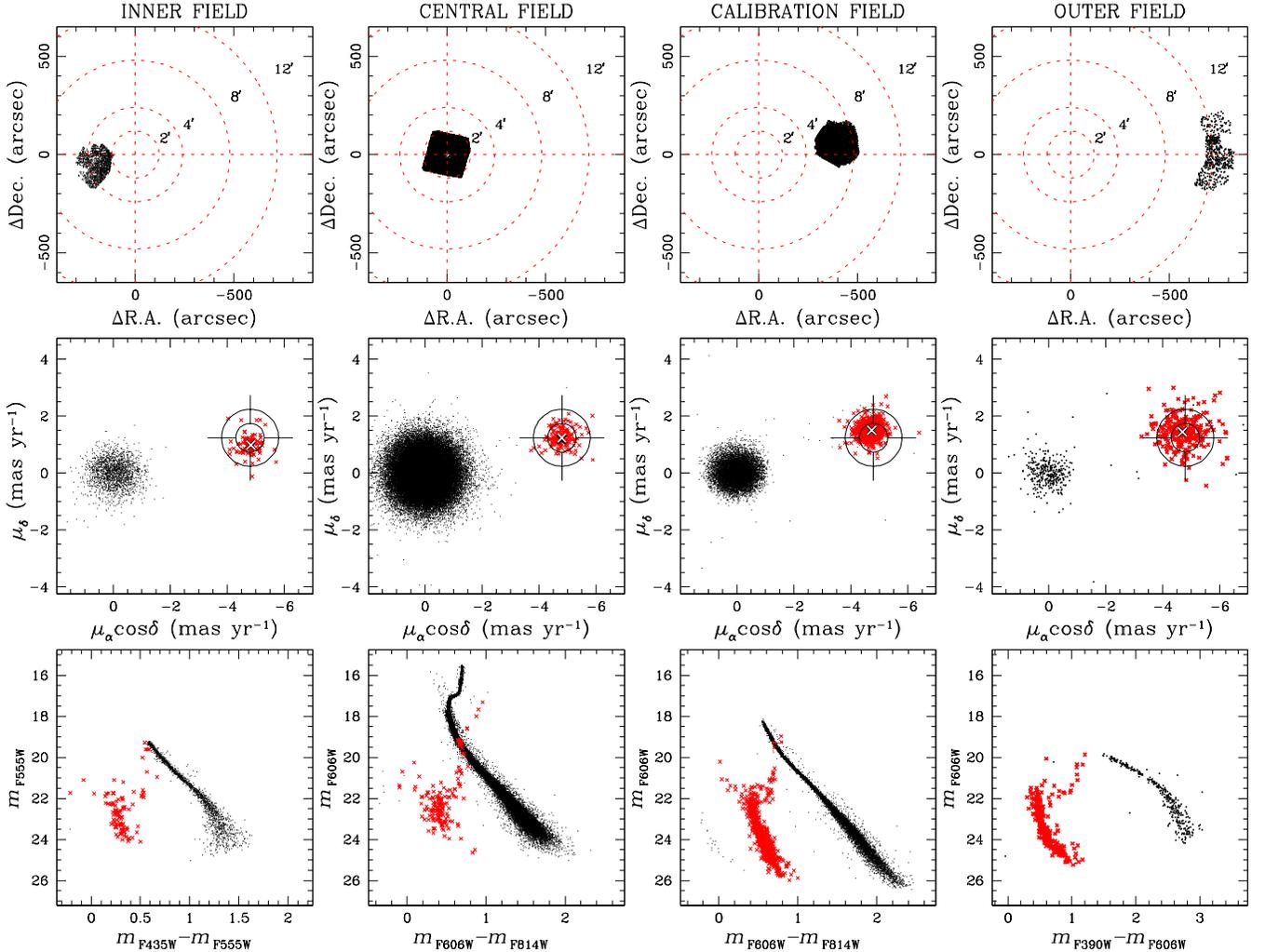}\\
\caption{From left to right:\ the inner field, the central field, the
  calibration field, and the outer field. From top to bottom, for each
  field:\ the field-of-view, the PM diagram, and the CMD. SMC stars are
  highlighted in red in both the PM diagrams and the CMDs. Concentric
  circles in the top panels, in red, give an idea of the radial
  extension of the data.  The crosshairs in each of the PM diagrams
  are centered on the barycenter of SMC stars in the central field.
  We can already see that SMC stars in the inner field (to the east
  side of the cluster's center) have preferentially lower $\mu_\delta$
  values, while the opposite happens to the fields on the west-side of
  the cluster's center. This is a clear sign of rotation. Moreover, we
  can see that, while the PM distribution of 47Tuc stars is circular
  in the central field, it is flatter in the inner and calibration
  fields; this is a clear sign of anisotropy. See the text for more
  details.\\~\\}
\label{f1}
\end{figure*}

\begin{figure*}[ht!]
\centering
\includegraphics[width=16cm]{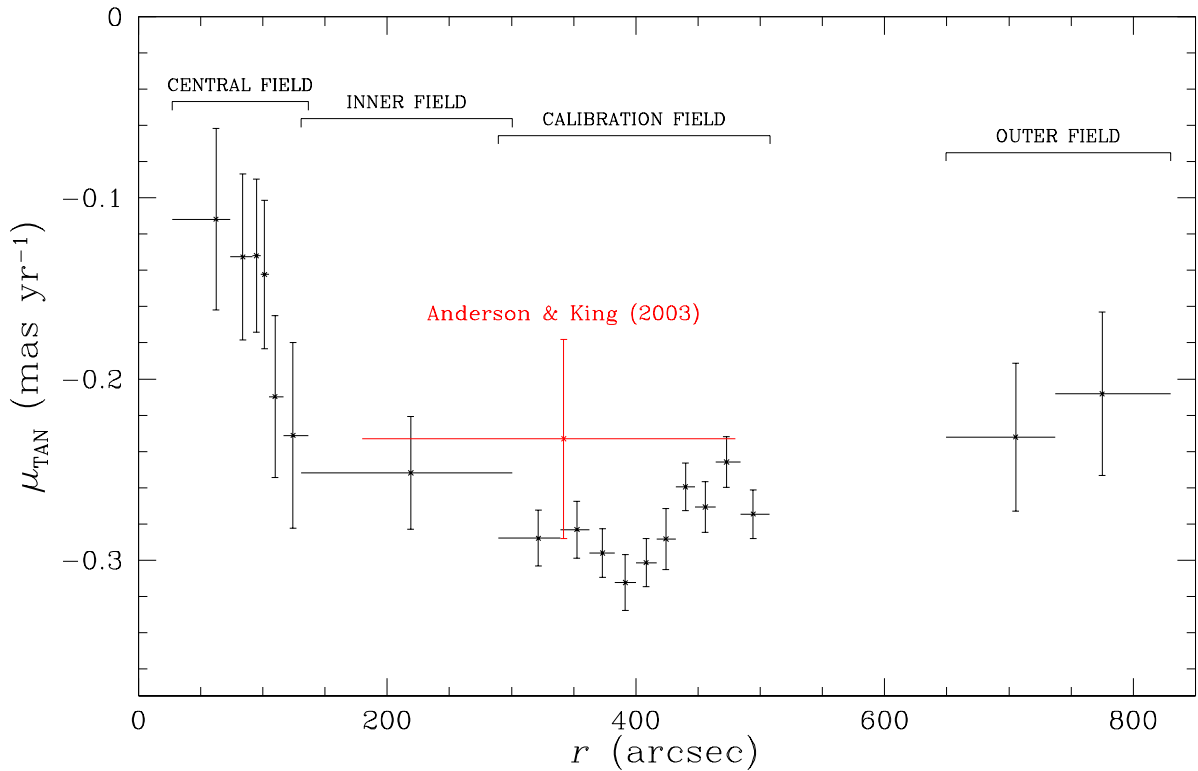}\\
\caption{Rotation of 47~Tuc ($\mut$) in the plane of the sky measured
  in the four fields, with errors. For completeness, we included the
  value computed by \citet{ak47t}. The horizontal bars indicate the
  radial intervals over which each $\mut$ value is determined.\\~\\}
\label{f2}
\end{figure*}

\begin{figure}[ht!]
\centering
\includegraphics[width=\columnwidth]{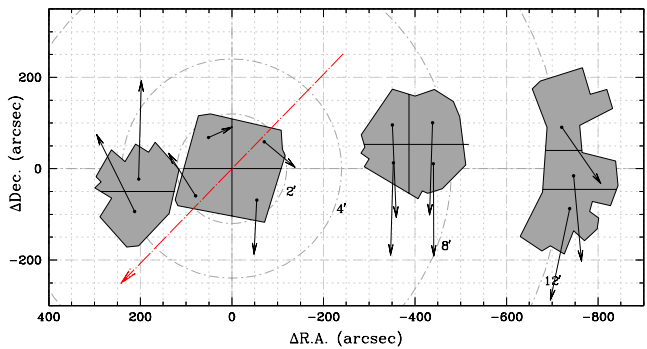}\\
\caption{The rotation map of 47~Tuc in the plane of the sky. Vectors
  are magnified by a factor 750. For completeness, we also show the
  LOS-based rotation axis (136$^\circ$ North to East,
  \citealt{paolo13}).\\~\\}
\label{f3}
\end{figure}

\subsection{Inner and outer fields}
\label{ss:data:wfpc2}

The inner and outer fields were observed with either ACS/WFC or
WFC3/UVIS in only one epoch.  All other available observations were
taken with the WFPC2. The inner field consists of one single ACS
pointing and all the available WFPC2 exposures that overlap it. The
outer field consists of three marginally-overlapping UVIS pointings,
roughly aligned along the Declination direction, and all the available
WFPC2 exposures overlapping them. The complete list of observations
for the inner and outer fields are reported in Tables~2 and 3,
respectively.

The PM-reduction tools presented in Paper~I were not designed to be
used on WFPC2 data, because the astrometric precision reachable with
WFPC2 exposures is generally far less that what we can achieve with
later \textit{HST} optical imagers.  The WFPC2 camera was composed of
four 800$\times$800 pixel$^2$ 12-bit detectors, with a pixel scale of
99.6 mas pixel$^{-1}$ for the three WF chips, and 45.5 mas
pixel$^{-1}$ for the PC chip.  All four chips suffered from $\sim$16\%
vignetting.  The incident light was split into four beams (one per
chip) by a four-faced pyramid mirror, which provided a challenging
registration of the observed relative chip positions. Finally, there
is no CTE correction available for the WFPC2.

Nevertheless, WFPC2 observations represent the only available first
epochs for the inner and outer fields, and allowed us to compute PMs
with a time baseline of 18.6 and 8 years, respectively. Our
PM-reduction tools are scalable, and it is straightforward to include
data coming from different instruments/cameras. We treated each WFPC2
chip independently, so to minimize inter-chip transformation
errors. We modeled single-exposure expected astrometric errors as a
function of the instrumental magnitude using all the available WFPC2
exposures in the two fields. The expected errors are used as a
first-guess weight during the PM-fitting procedures (see Sect.~5.2 of
Paper~I for more details). 

The computed PMs for the inner and the outer fields are in
  agreement with with those computed in the central and calibration
fields in terms of expected intrinsic values and associated
  errors.  On the other hand, the available dataset does not allow us
to adequately study and minimize the impact of systematic effects in
the quoted PM errors, as we did for the ACS and UVIS detectors. As a
result, we cannot as reliably study the internal kinematics of 47~Tuc
stars (which relies the subtraction in quadrature of PM errors) using
inner- and outer-field measurements as for the central and
  calibration field.

The inner-field PM catalog contains 2187 objects, 2084 of which passed
our high-quality selection criteria. The outer field consists of 648
total objects, of which 579 are identified as high-quality
measurements.

\section{The rotation of 47~Tuc in the plane of the sky}
\label{sec:rot}

The top panels of Figure~\ref{f1} show the field-of-view of each of the four
fields around the center of the cluster. From left to right we
have:\ (1) the inner field; (2) the central field; (3) the calibration
field; and (4) the outer field. Concentric circles, in red, give an
idea of the radial extension of the data.

\textit{HST} exposures all have different roll angles for different
epochs, and axis rotation is one of the six parameters that are solved
for when we transform stellar positions as measured on single
exposures into the master frame. Because of this, any direct sign of
cluster rotation (if present) is absorbed by the linear terms of the
coordinate transformations. Since our PMs are relative to the bulk
motion of the cluster, any other object in the field that is not a
cluster member would have a systematic component in its PM measurement
that is equal in size and with the opposite sign of the bulk rotation
of the cluster.

This systematic PM component is clearly visible in the PM diagrams
shown in the middle panel of Fig.~\ref{f1}. Cluster stars are in
black, while background SMC stars are marked as red crosses. As a
reference, a black crosshair in each panel highlights the barycenter
of SMC stars as measured in the central field where, because of
symmetry, the systematic PM rotation component is minimized. The white
crosses near the center of the crosshairs mark the median loci of SMC
stars in each of the four fields. The mean PM of SMC stars in the
inner field is shifted toward smaller $\mu_\delta$ values, while that
of the calibration and outer fields is shifted toward larger values of
$\mu_\delta$.  It is clear from the figure that SMC stars appear to
rotate counter-clockwise with respect to the center of 47~Tuc, which
directly translates into a clockwise rotation of 47~Tuc in the plane
of the sky.

\begin{figure*}[t!]
\centering
\includegraphics[width=15.5cm]{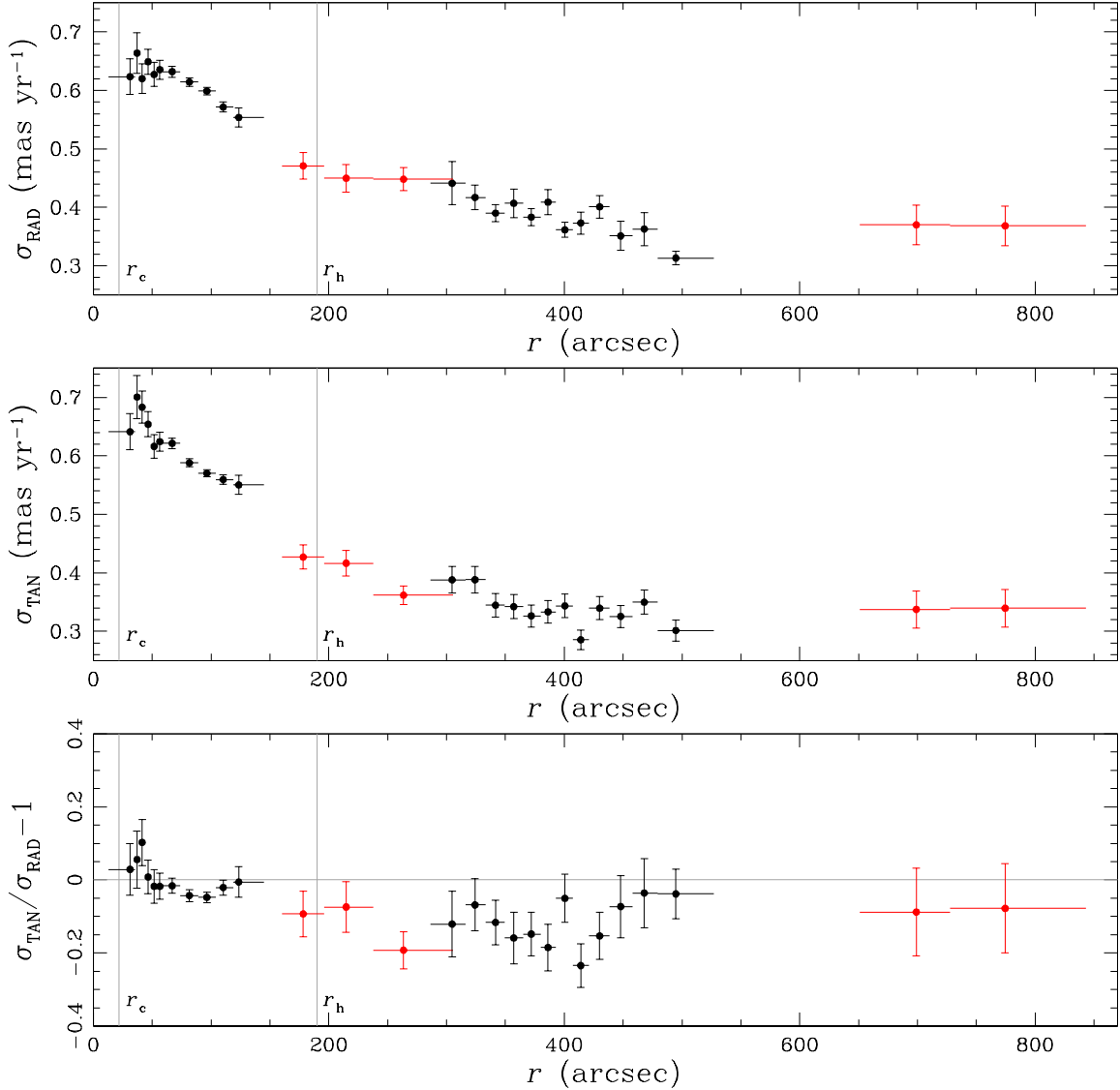}\\
\caption{Top panel:\ radial velocity-dispersion profile $\sigma_{\rm
    RAD}$ of MS stars in 47~Tuc as a function of distance from the
  cluster center. For completeness, we also show in red the values
  computed using WFPC2 data, for which systematic effects cannot be
  quantified. Middle panel:\ tangential velocity-dispersion profile
  for the same stars in the same radial bins. Bottom panel:\ deviation
  from isotropy (grey horizontal line). These profiles are based on MS
  stars in the magnitude range 20$<$$m_{\rm F606W}$$<$22,
  corresponding to a mass range between 0.62 and 0.47 $M_\odot$.  Core
  and half-light radii (\citealt{h96} values) are marked by
  the two vertical lines in all panels.\\~\\}
\label{f4}
\end{figure*}

The bottom panels of Fig.~\ref{f1} show the color-magnitude diagrams
(CMDs) of 47~Tuc stars (black dots) and SMC stars (red crosses) in
different magnitude/color combinations for the four fields. To select
bona-fide SMC stars we took advantage of their location on both the PM
diagrams and the CMDs. SMC stars on the PM diagram are clustered
around a well-defined location, clearly distinct from that occupied by
47~Tuc stars. We preliminary selected SMC candidates as all the
sources 4 sigma outside the distribution of 47~Tuc stars.  This rough
selection necessarily includes field stars and a few cluster
members. We then iteratively computed 4$\sigma$-clipped median values
$\mua$ and $\mud$ of the barycenter of SMC stars in the PM diagram,
and kept only stars within 4 $\sigma$ within this location. Finally,
we excluded by-eye from the SMC selection all those stars that did not
lie in the SMC region on the CMDs. We repeated one more time the
process of computing the location of the barycenter of SMC stars in
the PM diagram, and kept only SMC stars within 4 $\sigma$ from their
median location.

The final sample of SMC stars is what we show in red in
Fig.~\ref{f1}. We have, respectively:\ 96 SMC stars in the inner
field, 164 in the central field, 1495 in the calibration field, and
235 in the outer field.

In order to properly measure the apparent rotation of SMC stars, we
need to define a reference point (the zero point) on the PM diagram
with respect to which to compute their tangential component of the
motion ($\mut$).  To do this, we selected all SMC stars within the
largest circle that can be fully contained within the central field,
and determined a 3$\sigma$-clipped estimate of their barycenter
location on the PM diagram. This location has
coordinates:\ $\mua=-4.797$ $\masyr$, $\mud=1.244$ $\masyr$, and it
defines the center of the crosshairs in the middle panels of
Fig.~\ref{f1}.

We zero-pointed the PM of SMC stars to the reference point defined
above, and computed their average $\mut$ at different
equally-populated radial intervals with respect to the cluster's
center. A finer radial subdivision (about 28 stars per bin) is applied
to the central field, where the fastest rotational variation is found
(at the cost of larger measurement errors).  We then derived a single
rotation measurement for the inner field (96 stars), 10 measurements
for the calibration field (about 150 stars each), and 2 measurements
for the outer field (117 stars each). The computed quantities, as a
function of the radial distance, are shown with errorbars in
Fig.~\ref{f2}, and are listed in Table~4.  We adopted the convention
of using negative $\mut$ values in case of a clockwise rotation in the
plane of the sky (North to West). The horizontal bars in Fig.~\ref{f2}
indicate the size of the radial intervals within which each $\mut$
value is computed. For completeness, we included (in red) the rotation
value computed by \citet{ak47t}. All these quantities are listed in
Table~4.  The cluster has a solid-body-like rotation within our
central field. A simple least-squares fit of the form $\mut=m r$ gives
$m=-1.65\pm0.10$ $\masyr$ arcsec$^{-1}$. Then, the rotation profile
slowly flatters at larger radii, to reach the highest value of
$-0.312\pm0.015$ $\masyr$ at $\sim 390^{\prime\prime}$ ($\sim
6\farcm52$) from the cluster center. The measured rotation slows down
to about $-0.21$ $\masyr$ in the outermost regions probed by our data.

The general shape of the rotation curve of 47~Tuc we have measured on
the plane of the sky is qualitatively similar to that obtained by
\citet{paolo13} using LOS measurements. There is,
though, an important difference between the two rotation
profiles:\ the PM-based profile has a rotation peak in the plane of
the sky that is about twice as large as that measured with LOS
velocities, which we model in Sect.~\ref{sec:theo}.

To give the reader a better sense of how the cluster is rotating on
the plane of the sky, we show in Fig.~\ref{f3} the rotation map of the
cluster derived with the datasets we analyzed. To obtain the map, we
divided the datasets into either four quadrants (for the central
field) or equally-populated regions (for the other three fields), and
computed the zero-pointed median $\mua$,$\mud$ components of the
motion of SMC stars in each region. These values (with the opposite
sign) are shown as vectors departing from the median location of SMC
stars within each region. The length of the vectors, in $\masyr$, has
been magnified by a factor 750, for clarity. For completeness, we show
in red the rotation axis as measured from LOS data (136$^\circ$ North
to East, \citealt{paolo13}).

\section{Velocity-dispersion and anisotropy profiles}
\label{s:vdap}

The careful reader might have noticed that the distribution of 47~Tuc
stars in the PM diagram of the central field is rather circular, while
this distribution is more flattened in the inner and calibration
field, to become again somewhat more circular in the outer field
(middle panels of Fig.~\ref{f1}). This is the effect of
velocity-dispersion anisotropy.

To properly measure the degree of velocity-dispersion anisotropy of
the cluster as a function of radius, we proceeded as follows.  First,
we need to select stars with reliable PMs and similar masses in the
four fields.  As we mentioned earlier, PMs computed in the inner and
outer fields are based on WFPC2 measurements, and might be affected by
significant systematic effects. Our datasets do not allow us to
adequately study and minimize the impact of systematic effects due to
WFPC2 measurements. While these systematic errors are expected to only
have second-order effects in the quoted PMs, they can significantly
alter the estimated PM errors. Since PM errors are subtracted in
quadrature when we want to compute velocity-dispersion profiles,
under/overestimating PM errors could lead to incorrect profiles.  In
what follows, we will include measurements coming from the inner and
the outer fields only, for completeness.

To select stars of similar mass in the four fields, we limit our
selections to the magnitude range 20$<$$m_{\rm F606W}$$<$22. The
radial and tangential velocity-dispersion profiles were estimated
using the same method as in \cite{2010ApJ...710.1063V}, which corrects
the observed scatter for the individual PM uncertainties.

The top panel of Fig.~\ref{f4} shows the radial velocity-dispersion
profile $\sigma_{\rm RAD}$ of 47~Tuc as a function of radius. In red
we report the values computed in the inner and outer field, for
completeness. The horizontal bars illustrate the radial extent over
which each point is obtained. The central radial velocity dispersion
is about 0.64 $\masyr$, or 13.65 $\kms$ at a distance of 4.5 kpc (to
be compared to 11.5 $\kms$ for evolved stars,
\citealt{h96}).\footnote[1]{We are assuming here that the
    LOS-based velocity-dispersion-profile value quoted in the Harris
    catalog, which is the average of available measurements in the
    literature, refers to the center of the cluster. This might
    actually not be true, as literature values are computed over
    different radial ranges. We will see later, in
    Section~\ref{correction}, that our $\sigma$-mass dependence
    modeling predicts a red-giant-branch- (RGB-) mass scaled central
    velocity-dispersion value closer to about 12.5 $\kms$ rather than
    11.5 $\kms$.} The profile reaches a minimum value of about 0.31
$\masyr$ in the outermost point of the calibration-field
($\sim$$8\farcm25$, or about 495$^{\prime\prime}$).  The middle panel
of the figure shows the tangential velocity-dispersion profile
$\sigma_{\rm TAN}$ computed in the same radial intervals as for
$\sigma_{\rm RAD}$. The two vertical lines mark the location of the
core radius $r_{\rm c}$=$0\farcm36$ ($21\farcs6$) and the half-light
radius $r_{\rm h}$=$3\farcm17$ ($190\farcs2$)
(\citealt{h96}). Table~5 lists the values of the data points
  shown in Fig.~\ref{f4}.

The deviation from isotropy ($\sigma_{\rm TAN}$/$\sigma_{\rm RAD}$-1)
as a function of radius is shown in the bottom panel of
Fig.~\ref{f4}. The horizontal line at 0 indicates an isotropic
system. The center of the cluster is isotropic, with increasing radial
anisotropy moving outward. It is worth noting that this trend agrees
with what we saw in Paper~II for the entire sample of 22 GCs, albeit
over a smaller radial range.  The degree of anisotropy is significant
in the calibration field, i.e. at radial distances between 5$^\prime$
(300$^{\prime\prime}$) and 8$^\prime$ (480$^{\prime\prime}$).

Both the radial and the tangential velocity-dispersion profiles of
47~Tuc seem to drop in the centermost radial bin. The first and the
second points of the dispersion profiles are nonetheless consistent
with each other to within less than one sigma. Changing the size of
the first few radial bins produced similar trends, all consistent with
being flat in the centermost regions within the errorbars. Also note
that these profiles are based on relatively faint MS stars, which
suffer the most from the highly-crowded conditions of the core of the
cluster.  The velocity-dispersion profile of 47~Tuc we published in
Paper~II, based on much brighter, higher S/N RGB stars, does not show
any central drop.

\begin{table}[t!]
\centering
\scriptsize{
\begin{tabular}{cccc||cccc}
\multicolumn{8}{c}{\textbf{\small Table~4}}\\
\multicolumn{8}{c}{\textsc{\small Rotation of 47~Tuc in the Plane of the Sky}}\\
\hline\hline
$\langle r \rangle$ & $\!\!$N$_{\rm STAR}$$\!\!$ &  $\langle \mu_{\rm TAN}\rangle$ & $\langle \sigma_{\mu_{\rm TAN}}\rangle$&$\langle r \rangle$ &  $\!\!$N$_{\rm STAR}$$\!\!$  &$\langle\mu_{\rm TAN}\rangle$ & $\langle \sigma_{\mu_{\rm TAN}}\rangle$\\
($^{\prime\prime}$)  &   &$\!\!$($masyr$)$\!\!$  & $\!\!$($\masyr$)$\!\!$&($^{\prime\prime}$)  &   &$\!\!$($\masyr$)$\!\!$  & $\!\!$($\masyr$)$\!\!$\\
\hline
   62.39 & 28 &  $-$0.112 &      0.050                            &  373.12 & 150 &  $-$0.296 &      0.013\\  
   83.97 & 28 &  $-$0.132 &      0.046                            &  391.51 & 150 &  $-$0.312 &      0.015\\  
   94.92 & 28 &  $-$0.132 &      0.042                            &  408.38 & 150 &  $-$0.301 &      0.013\\  
  101.42 & 28 &  $-$0.142 &      0.041                            &  424.24 & 150 &  $-$0.288 &      0.017\\  
  109.86 & 28 &  $-$0.210 &      0.045                            &  440.00 & 150 &  $-$0.260 &      0.013\\  
  124.28 & 24 &  $-$0.231 &      0.051                            &  455.87 & 150 &  $-$0.271 &      0.014\\  
  219.02 & 96 &  $-$0.252 &      0.031                            &  472.92 & 150 &  $-$0.246 &      0.014\\  
  321.49 & 150&  $-$0.288 &      0.015                            &  494.26 & 145 &  $-$0.275 &      0.013\\  
 (342)$^{\dagger}$&&  ($-$0.233)$^{\dagger}$  &    (0.055)$^{\dagger}$   &  705.53 & 118 &  $-$0.232 &      0.041\\  
  352.61 & 150&  $-$0.283 &      0.016                            &  775.05 & 117 &  $-$0.208 &      0.045\\  
\hline\hline
\multicolumn{8}{l}{\footnotesize{${\dagger}$ From \citet{ak47t}.}}\\
\end{tabular}}
\end{table}

\begin{table}[t!]
\centering
\scriptsize{
\begin{tabular}{cccccc}
\multicolumn{6}{c}{\textbf{\small Table~5}}\\
\multicolumn{6}{c}{\textsc{\small Radial and tangential velocity-dispersion}}\\
\multicolumn{6}{c}{\textsc{\small profile data of 47~Tuc}}\\
\hline\hline
$\langle r \rangle$ & N$_{\rm STAR}$ &  $\langle \sigma_{\rm RAD}\rangle$ & err$_{\sigma_{\rm RAD}}$ & $\langle\sigma_{\rm TAN}\rangle$ & err$_{\sigma_{\rm TAN}}$\\
($^{\prime\prime}$)  &   &($\masyr$)  & ($\masyr$)  &($\masyr$)  & ($\masyr$)\\
\hline
\multicolumn{6}{c}{Central field}\\
\hline
  31.15 &  210  &   13.39 &  0.64        &      13.68 &  0.66 \\
  37.15 &  209  &   14.16 &  0.74        &      14.94 &  0.78 \\
  41.45 &  308  &   13.22 &  0.54        &      14.58 &  0.59 \\
  46.53 &  461  &   13.84 &  0.45        &      13.95 &  0.45 \\
  51.64 &  540  &   13.38 &  0.44        &      13.15 &  0.43 \\
  56.52 &  675  &   13.55 &  0.35        &      13.32 &  0.34 \\
  66.95 & 2589  &   13.48 &  0.20        &      13.26 &  0.20 \\
  81.60 & 3305  &   13.11 &  0.17        &      12.54 &  0.15 \\
  96.49 & 3588  &   12.77 &  0.13        &      12.16 &  0.13 \\
 110.30 & 2160  &   12.19 &  0.18        &      11.93 &  0.18 \\
 123.46 &  671  &   11.81 &  0.35        &      11.74 &  0.35 \\
 134.10 &  143  &   12.40 &  0.87        &      11.53 &  0.81 \\
\hline
\multicolumn{6}{c}{Inner field}\\         
\hline
 178.30 &  220 &  10.04 &  0.49          &       9.11 &  0.44\\
 214.73 &  220 &   9.59 &  0.50          &       8.88 &  0.47\\
 263.52 &  213 &   9.56 &  0.42          &       7.72 &  0.34\\
\hline
\multicolumn{6}{c}{Calibration field}\\                         
\hline
 304.81 &   147 &    9.41 &  0.79        &       8.27 &  0.48\\
 324.16 &   147 &    8.89 &  0.44        &       8.28 &  0.48\\
 341.59 &   147 &    8.32 &  0.31        &       7.35 &  0.43\\
 357.14 &   147 &    8.68 &  0.52        &       7.30 &  0.43\\
 372.00 &   147 &    8.17 &  0.32        &       6.96 &  0.40\\
 386.25 &   147 &    8.72 &  0.45        &       7.11 &  0.42\\
 400.55 &   147 &    7.71 &  0.28        &       7.32 &  0.43\\
 413.98 &   147 &    7.96 &  0.40        &       6.09 &  0.36\\
 430.15 &   147 &    8.55 &  0.41        &       7.24 &  0.42\\
 448.01 &   147 &    7.49 &  0.53        &       6.94 &  0.41\\
 468.00 &   147 &    7.74 &  0.60        &       7.46 &  0.44\\
 494.83 &   138 &    6.68 &  0.25        &       6.43 &  0.39\\
\hline
\multicolumn{6}{c}{Outer field}\\                               
\hline
 699.00 &   78  &   7.89  & 0.73         &       7.20 &  0.68\\
 774.47 &   77  &   7.86  & 0.73         &       7.25 &  0.68\\
\hline\hline
\end{tabular}}
\end{table}

\begin{figure*}[ht!]
\centering
\includegraphics[width=\textwidth]{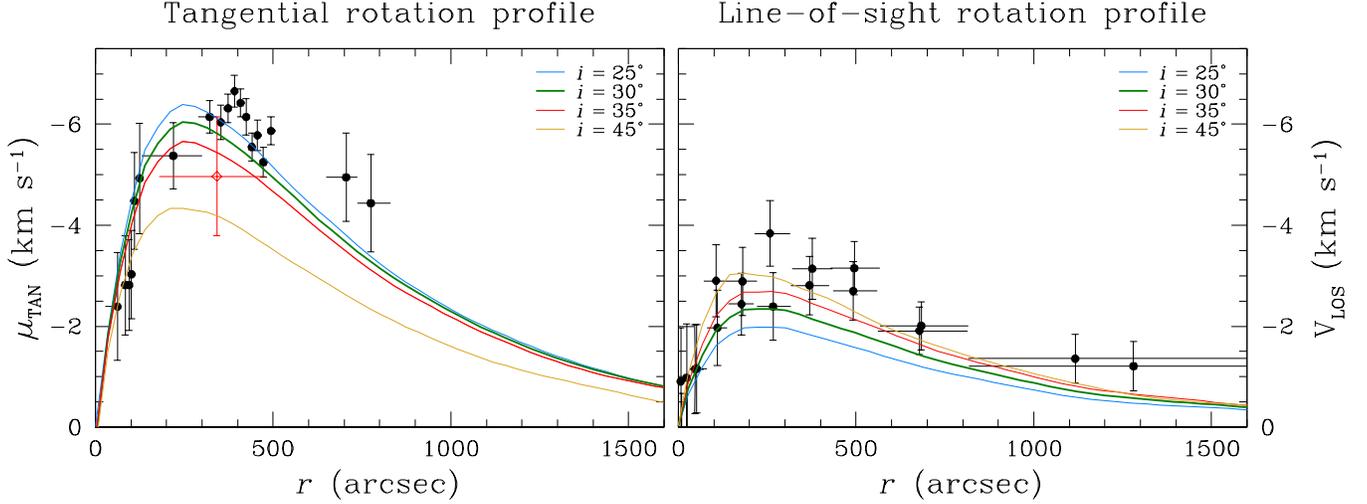}\\
\caption{Rotation profiles in the plane of the sky (from this work,
  expressed here in km\,s$^{-1}$ by assuming a cluster distance of 4.5
  kpc) and along the LOS (from \citealp{paolo13}), compared to the
  fits of the axisymmetric rotating models to all the
  available kinematic and photometric data. (The best-fit
    model provides an inclination angle $i=30^\circ$.)  The red
  point in the left panel refers to the measurement of \citet{ak47t}.
  The two profiles are shown here on in a similar representation, for
  clarity:\ the V$_{\rm LOS}$ profile was folded under the assumption
  of antisymmetry with respect to the cluster.  Blue, green, and red
  lines indicate models with inclination angles of $i=25^\circ$,
  30$^\circ$ and $35^\circ$, respectively. The yellow line indicates
  the model obtained by \cite{paolo13} assuming an inclination angle
  of $i=45^\circ$, clearly unable to reproduce the PM rotation
  profile. Our new model is able to much better reproduce
    simultaneously the three-dimensional rotation pattern of
    47~Tuc.\\~\\}
\label{f5}
\end{figure*}

\begin{figure}[ht!]
\centering
\includegraphics[width=0.5\textwidth]{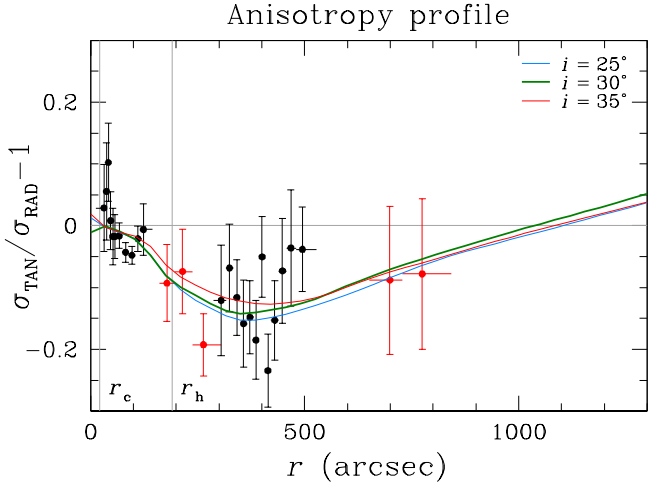}\\
\caption{Anisotropy profile predicted by our fitted models
  compared to the PM results from this work. Our model is able to
  reproduce well the observed trend of isotropy in the center and mild
  radial anisotropy in the outer parts.\\~\\}
\label{f6}
\end{figure}

\begin{figure}[ht!]
\centering
\includegraphics[width=0.5\textwidth]{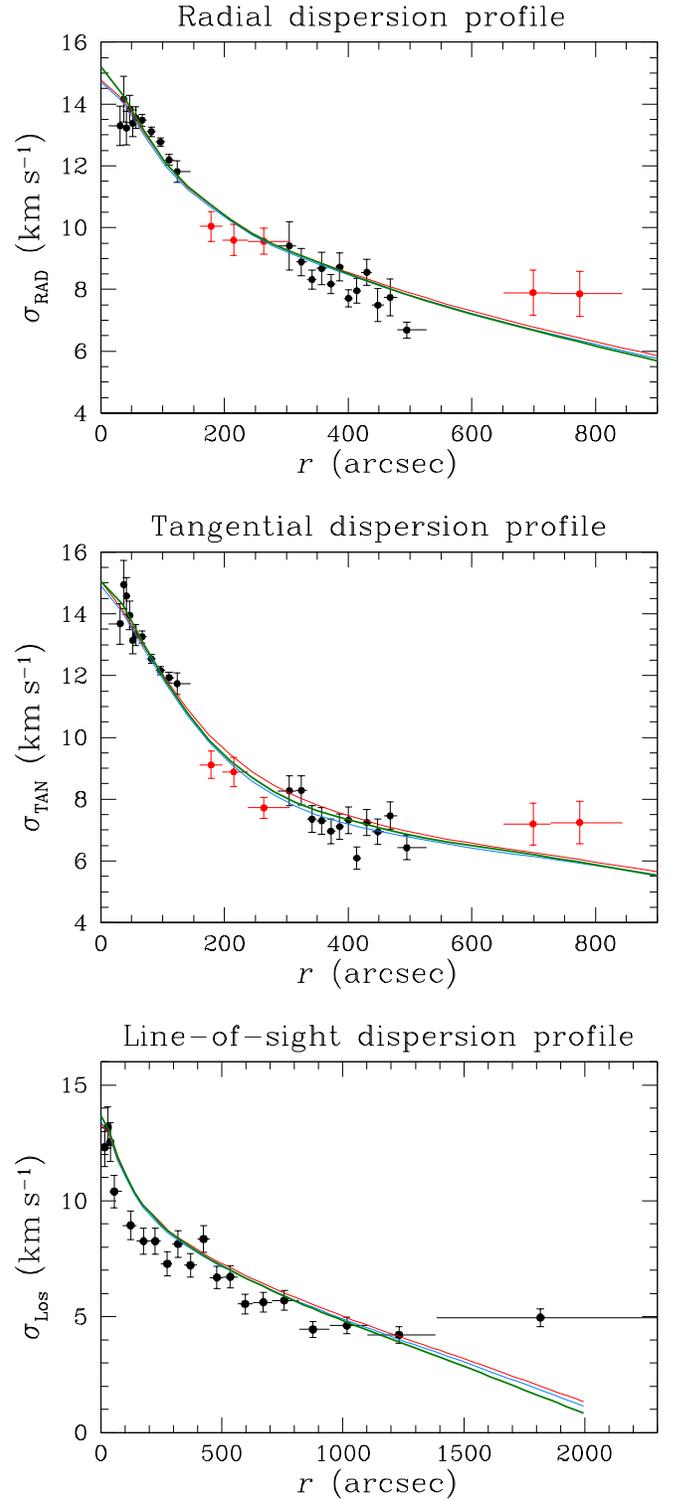}\\
\caption{Velocity-dispersion profiles along the tangential and radial
  components in the plane of the sky (from this work), and along the
  LOS (from \citealt{paolo13}), compared to our best-fit
  axisymmetric rotating model (in green). Note that the azure and the
  red lines are barely distinguishable from the green line.\\~\\}
\label{f7}
\end{figure}

\begin{figure}[ht!]
\centering
\includegraphics[width=0.5\textwidth]{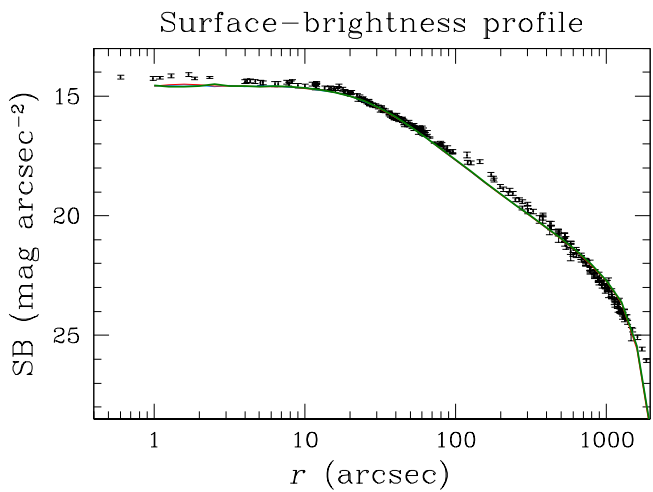}\\
\caption{Surface brightness profile \citep{Trager1995, 2006AJ....132..447N} compared to our
  best-fit axisymmetric rotating model (in green). Note that the azure
  and the red lines are barely distinguishable from the green line.\\~\\}
\label{f8}
\end{figure}

\begin{table*}[t]
\centering
\begin{tabular}{cccccccccc}
\multicolumn{10}{c}{\textbf{Table~6}}\\ 
\multicolumn{10}{c}{\textsc{Dimensionless Parameters and Physical Scales
    of the Best-Fit Model with $i=30^\circ$}}\\
\hline\hline
   &\multicolumn{3}{c}{Dimensionless parameters} & & \multicolumn{3}{c}{Physical scales} && Assumed distance\\
\cline{2-4}  \cline{6-8} \cline{10-10}
 & $\Psi$ & $\chi$ & $\bar{b}$ & & $\Sigma_0$ & $r_0$ & $v_0$ & &$d$ \\
 & & & & &mag arcsec$^{-2}$&arcsec&km s$^{-1}$& &kpc\\
\hline
 & 7.6 & $1.6\times10^{-3}$ & 0.007 &  & $14.5\pm0.1$  & $26.3\pm1.1$ & $14.3\pm0.1$ &  &$4.5$ \\
\hline\hline
\end{tabular}
\tablecomments{Concentration parameter $\Psi$, rotation strength
  parameter $\chi$, the $\bar{b}$ parameter (the additional parameter
  $c$ is set to unity), $V$-band central surface brightness $\Sigma_0$
  in mag arcsec$^{-2}$, radial scale $r_0$ in arcsec, velocity scale
  $v_0$ in km s$^{-1}$, and assumed distance d in kpc. For the
  physical scales, the associated $1\sigma$-errors are also
  shown.\\~\\}
\end{table*}

\begin{table*}[t]
\centering
\begin{tabular}{ccccccc}
\multicolumn{7}{c}{\textbf{Table~7}}\\
\multicolumn{7}{c}{\textsc{Structural Parameters Derived for the Best-Fit Model with $i=30^\circ$}}\\
\hline\hline
$C$ & $r_{\rm c}$ & $r_{\rm h}$& ${\rm R}_\mathrm{tr}$ & $M$ & $M/L_\mathrm{V}$& $\log\rho_0$ \\
&arcsec&arcsec&arcsec&$M_\odot$&$M_\odot/L_\odot$& $M_\odot$~ pc$^{-3}$\\
\hline
$1.91\pm0.02$	&$26.5\pm1.16$ & $240.7\pm10.6$& $1941.74\pm85.67$&$8.4\pm0.4\times10^5$ &$1.98\pm0.26$&$5.02\pm0.01$\\
\hline\hline
\end{tabular}
\tablecomments{Concentration parameter $C=~\log({\rm
    R}_\mathrm{tr}/r_{\rm c})$, projected core radius $r_{\rm c}$ in
  arcsec, projected half-light radius $r_{\rm h}$ in arcsec,
  tridimensional truncation radius ${\rm R}_\mathrm{tr}$ in cylindric
  coordinates (in arcsec), total mass of the cluster $M$ in units of
  $M_\odot$, V-band mass-to-light ratio in solar units, logarithm of
  the central mass density $\rho_0$ in units of $M_\odot$~
  pc$^{-3}$.\\~\\}
\end{table*}

\section{Dynamical models}
\label{sec:theo}

The goal of this section is to provide a dynamical model to describe
the comprehensive set of available observations, comprising the PM
kinematics analyzed above in addition to the classic LOS
data and photometry. We will use a family of physically-motivated
distribution-function based models (\citealp{VarriBertin2012}),
recently applied to a selected sample of Galactic GCs
(\citealp{Kacharov2014}, \citealp{paolo13}).

These self-consistent models have been specifically constructed to
describe quasi-relaxed stellar systems with realistic differential
rotation, axisymmetry and pressure anisotropy. The models are defined
by four dimensionless parameters (concentration parameter $\Psi$,
rotation strength parameter $\chi$, and the parameters $b$ and $c$
determining the shape of the rotation profile). A full description of
the distribution function and of the parameter space is given in
\citet{VarriBertin2012}. Since these models allow only for single-mass
component, they do not take into consideration the effects connected
with mass segregation. The implications for this assumption are
described in Sect.~\ref{correction}.

We will use the PM-based profiles described in Sect.~\ref{s:vdap}
(tangential and radial velocity dispersion profiles and rotation
profile in the plane of the sky), the LOS-based velocity
profiles (velocity dispersion and rotation profiles) reported in
\citet{paolo13}, and the photometric data from \citet{Trager1995}.
We do not consider in the fit procedure the data obtained from
  the inner and outer fields, as described in Sect.~\ref{s:vdap}.

\subsection{Fitting procedure}
\label{ss:fitproc}

We follow the same fitting procedure outlined in \citet{Kacharov2014}
and \citet{paolo13}, as summarized in the following. The comparison
between the differentially-rotating axisymmetric models and the
observations requires us to specify four dimensionless parameters and
five additional quantities:\ three physical scales (i.e., the radial
scale $r_0$, the central surface density $\Sigma_0$, and the velocity
scale $v_0$), the inclination angle $i$ between the rotation axis and
the LOS direction, and the cluster distance.  Since the
adopted dynamical models are characterized by deviations from isotropy
in configuration and velocity space, the choice of the inclination
angle plays a fundamental role in the fitting procedure. To exploit
such an additional degree of freedom, we initially assume the value
$i=30^{\circ}$, derived by applying Eqn.~8 of \cite{vandeVen2006}
(linking the average motion along the LOS to the one in the
plane of the sky) to our data. We will later explore how the best-fit
parameters change with different inclinations
($i=25^{\circ}$--$35^{\circ}$). Since PMs are measured in mas
yr$^{-1}$ in the plane of the sky, a multiplying factor of $4.74\,d$
is needed to convert these values to $\kms$. We will assume the
distance of 47 Tuc of $d=4.5$ kpc (\citealt{h96}).

The fitting procedure is twofold. First, we determine the dimensionless
parameters needed to reproduce the observed value of $V/\sigma$ and the
observed position of the rotation peak (for further details see
Sect.~3.1 and 3.5 of \citealt{paolo13}). Then we calculate the
physical scales through $\chi^2$ minimization for all the kinematic
profiles (three dispersion profiles, two rotation curves) to obtain
the radial scale $r_0$ and the velocity scale $v_0$, and then we fit
the surface-density profile to obtain the central surface density
$\Sigma_0$. This provides all the constrains needed to
determine the best-fit dynamical model.

We wish to emphasize that, during the fitting procedure, the
projection of the self-consistent dynamical models is performed by
sampling from the relevant distribution function a discrete set of
$N$=2\,048\,000 particles and then by performing a rotation of such a
discrete system to match the relevant inclination angle.  The
theoretical kinematic and photometric profiles are then calculated by
following the same procedures applied for the construction of the
observational profiles (i.e., by means of circular annuli in the
projection plane). Any emerging constraint on the morphology and
degree of anisotropy of the stellar system should be, therefore,
considered as resulting properties of the best-fit model, which has
been selected exclusively on the basis of the (spherically-averaged)
kinematic and photometric information.

As to the morphological characterization (see Sect.~\ref{ss:res}), the
projected isodensity contours are calculated on the basis of the
nonspherical projected number density distribution. The relevant
ellipticity profile is then constructed by considering the ratio of
the principal axes of approximately 60 isodensity contours,
corresponding to selected values of the normalized projected number
density in the range $[0.9, 10^{-3}]$; smooth profiles are
then obtained by performing an average on subsets made of
10--20 individual ellipticity values.

For completeness, we also explored the distance value of 4.15 kpc from
\citet[Paper~III]{2015ApJ...812..149W} in our fitting procedures. The
best-fit model based on the 4.15 kpc value provides comparable results
to those based on the 4.5 kpc distance value. However, the
Paper~III-based model offers a better fit to the LOS
velocity-dispersion profile, but a worse fit to the surface-brightness
profile.

\subsubsection{Correction for energy-equipartition effects}
\label{correction}

Since PM measurements sample different kinematic tracers than
LOS measurements --namely stars with lower mass than bright
red giant stars-- some caution is needed when applying a one-component
dynamical model simultaneously to the three-dimensional kinematics. In
particular, our PM-based velocity-dispersion profiles are constructed
using MS stars within the magnitude range 20$<$$m_{\rm
  F606W}$$<$22. This implies a typical stellar mass between 0.62 and
0.47 $M_\odot$ (using a \citealt{2008ApJS..178...89D} isochrone
with [Fe/H]=$-$0.5, [$\alpha$/Fe]=0.2, $d=4.5$ kpc and
  E($B$$-$$V$)=0.04). The LOS measurements have instead a
typical stellar mass of 0.83 $M_\odot$, since they sample only bright
giant stars.

Since GCs reach a state of partial energy equipartition (see e.g.,
\citealp{TrentivanderMarel2013}, \citealp{Bianchini2016}), their
kinematics are expected to show a mass dependence, with lower-mass
stars having gradually higher velocity
dispersions. \cite{Bianchini2016} showed that the mass-dependence of
kinematics depends on the relaxation condition of the cluster (see
their Eqn.~6). Therefore, by knowing the relaxation state of a cluster
it is possible to predict the shape of the velocity dispersion as a
function of mass $\sigma(m)$, and then rescale the PM-based
dispersion profiles according to the calculated factor.

Given $n_{\rm rel}=T_{\rm age}/T_{r_{\rm c}}=169.6$ (number of core
relaxation times the cluster has experienced), the mass dependence
of the velocity dispersion from Eqn.~3 of \cite{Bianchini2016} is
$$ 
\sigma(m)=\sigma_0\exp\left(-\frac{1}{2}\frac{m}{m_{\rm eq}}\right),
$$ with $m_{\rm eq}=1.60$ $M_\odot$ (from Eqn.~6 of
\citealp{Bianchini2016}) and $\sigma_0$ a normalization factor. Given
a mass of 0.83 $M_\odot$ for the LOS velocities and a median
value of 0.54 $M_\odot$ for the stars used for the PM-based
  velocity-dispersion profiles (estimated via the adopted isochrone),
the velocity dispersions of the two different mass-tracers are related
by $\sigma(0.83)/\sigma(0.54)=0.913$.  We rescale the tangential and
radial PM dispersion profiles of the model by dividing them by this
correction factor. We do not rescale the PM rotation profile, since we
do not observe any signature of dependence of mass.

Note that the central velocity-dispersion estimate based on
  the PM of evolved stars (Paper~II) is 0.573$\pm$0.005 $\masyr$,
  which translates into 12.2$\pm$0.1 $\kms$ at a distance of 4.5
  kpc. The adopted rescaling factor of 0.913 implies a central
  velocity-dispersion for RGB stars of 12.46 $\kms$, in full agreement
  with the value reported in Paper~II.

\begin{figure}[!t!]
\centering
\includegraphics[width=\columnwidth]{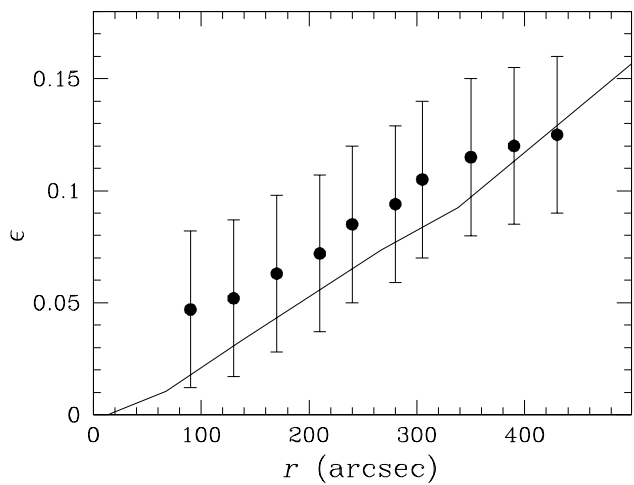}\\
\caption{Projected ellipticity profile for 47~Tuc. The black dots mark
  the observed ellipticity measured at different radii, as presented
  by \citet{1987ApJ...317..246W}; the solid line represents the
  profile derived from our best-fit axisymmetric rotating model.\\~\\}
\label{f10}
\end{figure}

\begin{figure}[!t!]  \centering
\includegraphics[width=\columnwidth]{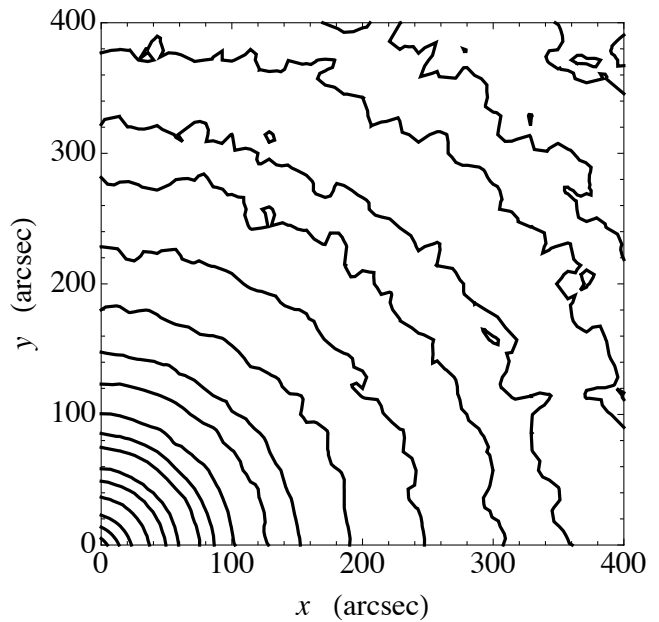}\\ 
\caption{Projected isodensity contours for the discrete
    realization of the best-fit axisymmetric model. The contours are
    calculated in the first quadrant of the projection plane and
    correspond to selected values of the projected number density
    (normalized to the central value) in the range $[0.9,
      10^{-3}]$. The area represented in the figure covers a square of
    side length approximately equal to $2$$\times$$r_{\rm h}$.\\~\\}
\label{f11}
\end{figure}

\begin{figure}[!t!]
\centering
\includegraphics[width=\columnwidth]{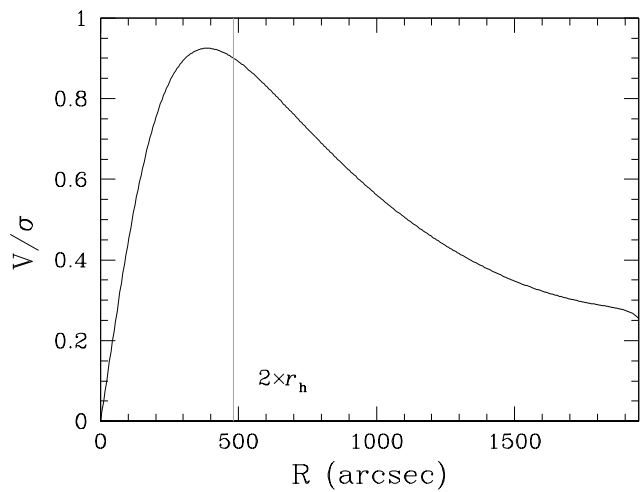}\\
\caption{The intrinsic V/$\sigma$ profile in the equatorial plane,
  based on the best-fit model. The profile reaches a peak value of
  $\sim$0.9 at around two half-light radii, indicating
  that 47~Tuc is rotating at a much higher rate than previously
  reported.\\~\\}
\label{f9}
\end{figure}

\begin{figure}[!t!]
\centering \includegraphics[width=\columnwidth]{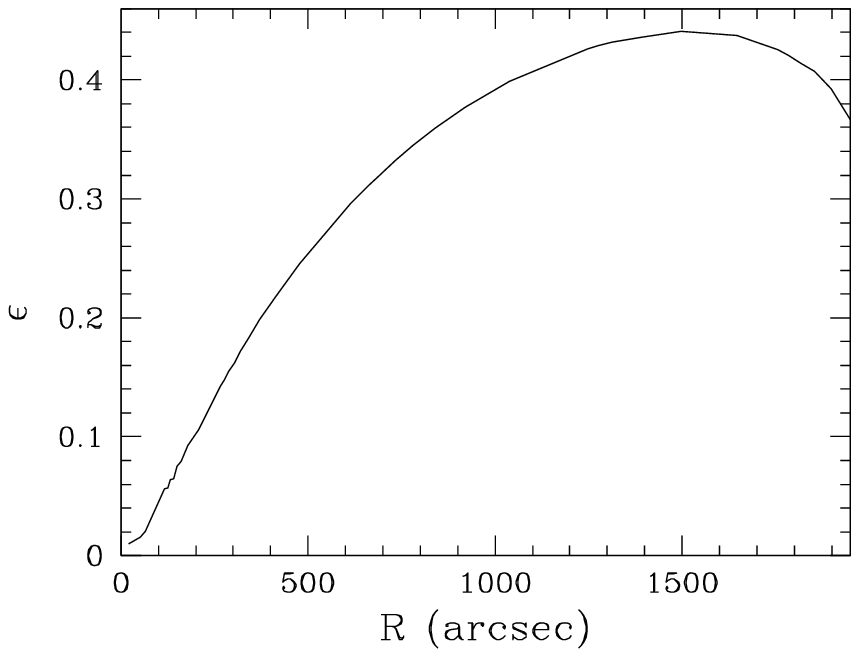}\\
\caption{Intrinsic ellipticity profile based on the best-fit
  model. The central region is characterized by an approximately
  linear behavior, as expected by the solid body-like behavior of the
  rotation curve in that radial range.  Some degree of flattening is
  still appreciable towards the boundary of the system, mostly as a
  result of the complex shape of the isodensity surfaces at low
  density.\\~\\}
\label{f12}
\end{figure}

\subsection{Results}
\label{ss:res}

\subsubsection{Projected properties}
\label{sss:proj_prop}

The results of the fit are reported in Figs.~\ref{f5}--\ref{f8}. The
azure, green and red lines in each plot correspond to the fit with
inclination angles $i=25$$^{\circ}$, 30$^{\circ}$ and 35$^{\circ}$,
respectively. The structural parameters of the three different models
do not differ significantly, however, the model with $i=30^{\circ}$
gives a better fit. We will consider this model as our fiducial
best-fit model and we report the best fit parameters and structural
properties in Tables~6 and 7.

Figure~\ref{f5} shows the rotation profile based in the LOS
direction (right) and in the plane of the sky (left). The additional
yellow lines represent the best-fit model derived by \cite{paolo13}
assuming an inclination of 45$^{\circ}$ and without accounting for the
rotation in the plane of the sky. The figure clearly shows that our
new best-fit model, assuming the new inclination value of
$i\simeq30^\circ$, is able to reproduce the three-dimensional
rotational structure of 47~Tuc, whereas a higher inclination angle
would fail in describing the rotation in the plane of the sky.

Our model predicts an anisotropy profile that is in excellent
agreement with the observations (Fig.~\ref{f6}), characterized by
isotropy in the central cluster regions, and mild radial anisotropy in
the intermediate regions. Moreover, we are able to simultaneously
reproduce all the dispersion profiles (Fig.~\ref{f7}) and the
surface-brightness profile (Fig.~\ref{f8}), employed for the fit.

Since the adopted dynamical equilibria are axisymmetric, we can
calculate the corresponding projected ellipticity profile, which we
illustrate in Fig.~\ref{f10}, together with the ellipticity data
currently available for 47~Tuc (from
\citealt{1987ApJ...317..246W}). We recall that the ellipticity profile
associated with the best-fit self-consistent model is a structural
property completely determined by the values of the dimensionless
parameters and physical scales identified during the model selection
procedure. As already appreciated in our previous analysis (see
\citealt{paolo13}), we find a very good agreement with the available
observational profile. Such a result allows us to confirm, with
increased confidence, that the physical origin of the observed
flattening of 47~Tuc is indeed the presence of internal rotation. We
emphasize that such an agreement is nontrivial, especially in
consideration of the reduced value of the inclination angle (in our
previous study we adopted $i=45^\circ$, while now we determined
$i=30^\circ$ to be more appropriate). The appreciable morphological
consistency recovered, once again, in the current analysis should be
interpreted as a manifestation of the more significant intrinsic
rotation (as recovered from the additional constraints posed by new PM
datasets), which compensates the effects of a less favorable
LOS direction.  For completeness, in Fig.~\ref{f11} we also
  show the projected isodensity contours of the discrete realization
  of the best-fit axisymmetric model as a function of the
  tridimensional radius in cylindrical coordinates.

\subsubsection{Intrinsic properties}
\label{sss:intr_prop}

Our three-dimensional model allows us to explore the intrinsic
kinematic and morphological structure of 47~Tuc.  The derived intrinsic
V/$\sigma$ profile, measured in the equatorial plane perpendicular to
the rotation axis, is characterized by a peak value of $\sim$0.9
reached at around 2 half-light radii (Fig.~\ref{f9}).  This confirms
that the internal rotation of 47~Tuc is higher than what has been
reported in previous studies (e.g., the LOS-based value
V/$\sigma$=0.25, \citealt{paolo13}). Correspondingly, we can also
characterize the three-dimensional structure of the cluster by means
of the intrinsic ellipticity profile, as measured in the meridional
plane of system (defined by the rotation axis and any of the principal
axes on the equatorial plane) and expressed as a function of the
semimajor axis (Fig.~\ref{f12}). The behavior of this profile in the
central regions is approximately linear, as shaped by the solid-body
like behavior of the rotation curve in that portion of the system. We
emphasize that the radial location of the peak of the ellipticity
profile does not correspond to the location of the peak of the
intrinsic $V/\sigma$ profile, as appropriate in the presence of a
nontrivial coupling between the angular momentum and mass distribution
within the system.

\begin{figure}[!t!]
\centering
\includegraphics[height=5.8cm]{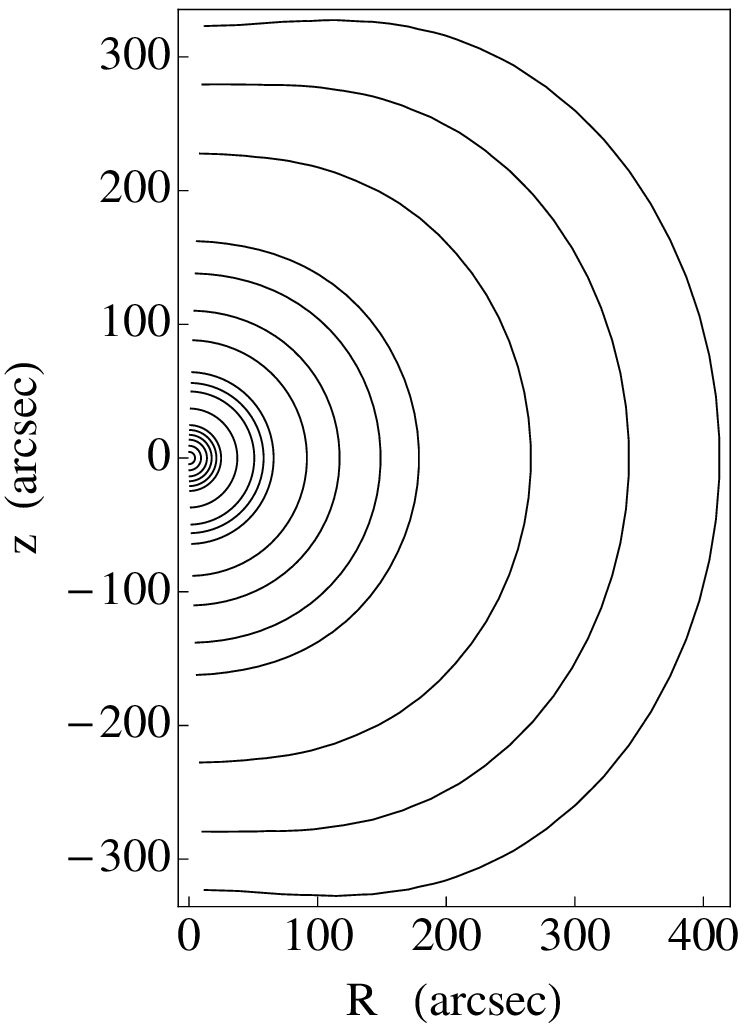}
\includegraphics[height=5.8cm]{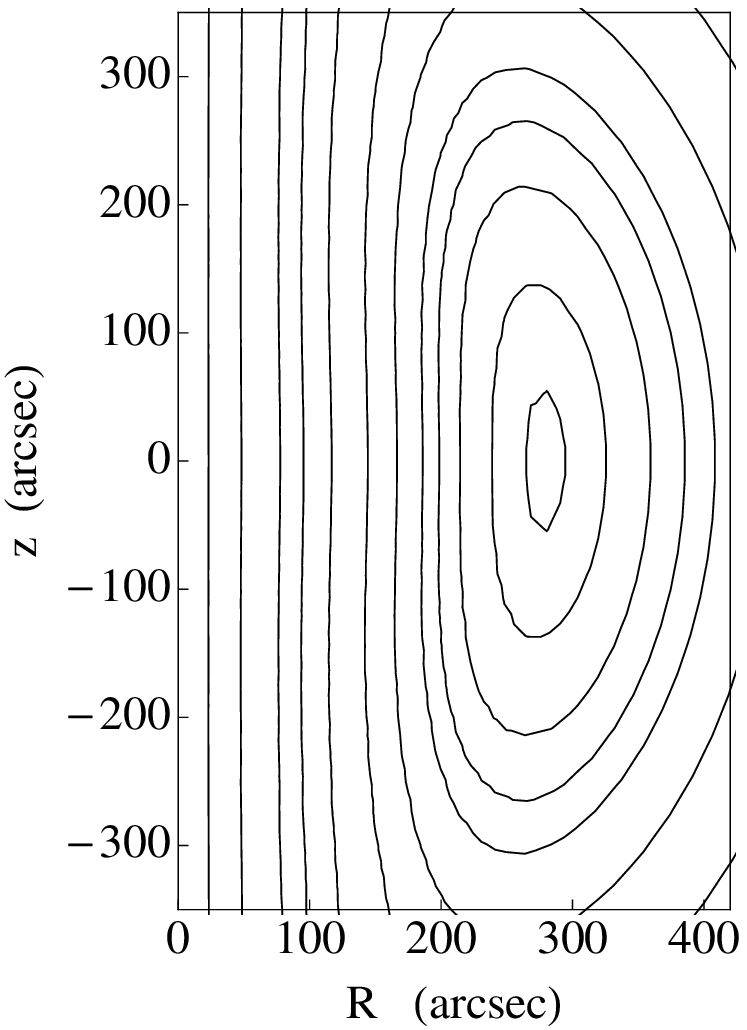}\\
\caption{Bidimensional maps of the intrinsic density and
  angular momentum distributions depicted as sections of the isodensity
  (left) and isovelocity (right) surfaces evaluated on the meriodional
  plane of the best-fit model (meridional sections). The vertical axis
  corresponds to the rotation axis, while the horizontal axis
  represents any principal axis on the equatorial plane. Note that the
  model is characterized by reflection symmetry with respect to such
  a plane.  \\~\\}
\label{f13}
\end{figure}

We recall that the intrinsic morphological and kinematic
  structure of an axisymmetric (nonstratified) rotating equilibrium is
  more complex than a simple characterization expressed by means of
  unidimensional radial profiles, therefore, we include also a
  representation of the bidimensional maps depicting the structure of
  the density and angular momentum distribution, expressed in terms of
  the contours of the normalized isodensity and
  isovelocity surfaces evaluated on the meridional plane of the
  three-dimensional model. In the maps illustrated in Fig.~\ref{f13}
  the radial range on the equatorial plane is approximately equal to
  $2r_{\rm h}$. From the density map (left panel) it may be
  appreciated that the maximum degree of flattening is reached in the
  intermediate regions of the system and that the shape of the
  isodensity contours becomes more complex as the density decreases
  (i.e., the outer contours are not monotonically increasing functions
  of the polar angle, hence the dimples on the rotation axis). From
  the kinematic map (right panel) it may be noted that the velocity
  field is not cylindrically-stratified, presenting an absolute
  maximum at about 300$^{\prime\prime}$; the location of the peak of
  the intrinsic $V/\sigma$ profile is determined by the interplay
  between the three-dimensional structure of such a velocity field and
  that of the trace of the velocity dispersion tensor. Equivalent maps
  may be calculated also for the meridional sections of the
  equipotential and isobaric surfaces.

Lastly, we wish to emphasize that the current presence of such an
appreciable degree of internal rotation in 47~Tuc carries important
implications about the initial amount of angular momentum of the
cluster. Indeed, many facets of the role played by internal rotation
during the long-term evolution of collisional stellar systems must
still be explored, but there is confirmed evidence that two-body
relaxation and mass loss determine transport and loss of angular
momentum in clusters (see, e.g., \citealt{1999MNRAS.302...81E,
  2007MNRAS.377..465E}), therefore any measurement of the presence of
internal rotation at the present day should be considered as a lower
limit of the initial angular momentum content. The emerging kinematic
complexity of 47~Tuc (and a progressively increasing number of
Galactic GCs) therefore offers an essential ingredient towards a more
complete and fundamental understanding of the formation and early
dynamical evolution of GCs.

\subsubsection{Some remarks on the modeling strategy}
\label{sss:remarks}

We recognize that our modeling strategy has a number of limitations,
especially regarding the description of the projected structural and
kinematic profiles in the outer regions of the cluster. With
particular reference to the behavior of the surface brightness profile
and LOS velocity dispersion profile in the very outer parts
(at radii $>$ 1500$^{\prime\prime}$), we interpret their discrepancies
as due to the effects of the tidal field of the Milky Way on the
structure and internal kinematics of the system.  It is well known
that, during the course of their dynamical evolution (as driven by
two-body relaxation), collisional stellar systems tend to expand until
they fill their Roche lobe, and then progressively start to lose
mass. Such a process significantly affects the phase-space
  structure of their periphery, both by increasing the complexity of
  the kinematics of the stars that are energetically bound (e.g.,
  \citealt{2016MNRAS.455.3693T,t2017}), and by determining the
existence of a population of energetically unbound stars that are
nonetheless spatially confined within the cluster (``potential
escapers'', see \citealt{2010MNRAS.407.2241K, 2017MNRAS.466.3937C}).
In terms of projected observables, these evolutionary effects may
manifest themselves in the form of surface-brightness profiles
extending beyond the cut-off radius predicted by a simple spherical
King model (i.e., the so-called ``extra-tidal'' structures identified
in many globular clusters in Local Group galaxies, see
e.g. \citealt{2005ApJS..161..304M, 2002AJ....123.1937B,
  2002AJ....124.1435H}) and velocity dispersion profiles characterized
by an untruncated, flattened behavior (e.g., see the studies of M15
and M92 by \citealt{1998AJ....115..708D, 2007AJ....133.1041D} and
NGC~5694 by \citealt{2015MNRAS.446.3130B}).  In term of intrinsic
properties, tidally-perturbed systems are also characterized by a
flavor and degree of anisotropy that strongly depends on the strength
of the tidal environment in which they have evolved (see
\citealt{2016MNRAS.461..402T} and additional remarks at the end of
next paragraph).

In this respect, we emphasize that simple, physically-based
  dynamical models of the kind adopted in our study, as defined by a
  quasi-Maxwellian distribution function, suitably modified near the
  tidal boundary and truncated above it, successfully capture the
  phase-space properties of the bulk of the cluster members by relying
  on a truncation prescription that heuristically mimic the effects of
  the tidal field (e.g. \citealt{1954MNRAS.114..191W,
    1966AJ.....71...64K, 1975AJ.....80..175W} and, more recently,
  \citealt{2015MNRAS.454..576G}).  Of course, the choice of such a
  one-dimensional (energy) truncation prescription in the definition
  of the distribution function strongly affects the structural and
  kinematic properties of the resulting configurations (see
  \citealt{1977A&A....61..391D, 1977AJ.....82..271H}), but, by
  definition, they can not offer a realistic description of the tidal
  field (see \citealt{1995MNRAS.272..317H, 2008ApJ...689.1005B}) or
  account for the existence of energetically unbound members (see
  \citealt{2017MNRAS.468.1453D}).  We wish to emphasize that, despite
  their inherent simplicity, the self-consistent rotating models
  adopted here have nonetheless two advantageous properties. First,
  their phase-space truncation reduces, in the non-rotating limit, to
  the smooth \citet{1975AJ.....80..175W} prescription, which is now
  often considered more successful in describing the outer regions of
  clusters than the traditional \citet{1966AJ.....71...64K} model (see
  \citealt{2005ApJS..161..304M}), which demands only continuity in
  phase space. Second, their velocity dispersion tensor is
  characterized by isotropy in the central region, weak radial
  anisotropy in the intermediate regions, and tangential anisotropy in
  the outer parts. Such a behavior of the pressure tensor was not
  assigned a priori in the definition of the models, but it results
  from the requirement of self-consistency, positivity of the
  distribution function, and energy truncation in the presence of
  angular momentum as a second integral of the motion. Isotropy or
  mild tangentially-biased pressure anisotropy in the outer parts of a
  star cluster has been shown to be a natural result of the dynamical
  evolution of a collisional stellar system within an external tidal
  field, which induces a preferential loss of stars on radial orbits
  (e.g., see \citealt{1997MNRAS.292..331T, 1997MNRAS.286..709G,
    2003MNRAS.340..227B, 2017MNRAS.466.3937C, 2016MNRAS.461..402T,
    2014MNRAS.443L..79V}).

The other discrepancy, namely a model underestimate of the
  surface brightness in the very central region and in an intermediate
  region in the range (100$^{\prime\prime}$--200$^{\prime\prime}$),
  can instead be considered as a result of a $\chi^2$-minimization
  employed by the fitting procedure that gives more weight to the
  three-dimensional kinematic data (see Sect.~\ref{ss:fitproc}). As a
  result, the value of the best-fit half-light radius we obtain
  ($r_{h} =240.7\pm10.9$ arcsec) is larger than what is reported in
  previous literature works (e.g., $r_{h}=190.2$
  arcsec,\citealt{h96}).

We wish to emphasize that our study focuses on understanding
  the rotational structure of 47~Tuc.
    The kinematic profiles (the two rotation curves and the three
    velocity dispersion profiles) show a general agreement within the
    observational error bars. However our model underestimates the
    values of the rotation profiles in the intermediate region and
    overestimates the value of velocity dispersion along the
    LOS. These discrepancies could be interpreted in view of
    the intrinsic limitations of the choice of the distribution
    function defining the dynamical models under
    consideration. Indeed, alternative modeling strategies may offer
    additional degrees of freedom in the parametrization of the phase
    space, but we wish to note that differentially rotating equilibria
    are relatively rare, especially for globular cluster studies (see
    \citealt{2009MNRAS.396.2183S} for an application of the rotating
    models proposed by Wilson 1975 to the study of $\omega$~Cen). An
    alternative approach for the construction of phase-space
    equilibria is based on the use of actions instead of integrals of
    the motion; an example of this line of attack has been recently
    proposed by \citet{2017arXiv170407833J}.  Finally, we stress that
    the velocity anisotropy profile and the ellipticity profile are
    not directly used in the fitting procedure and the respective
    figures show a comparison between the available data and the model
    prediction. This result may also be interpreted as a reassuring
    ``a posteriori'' validation of our choice of adopting axial
    symmetry and reduces the scope of exploring more general
    symmetries (e.g., triaxial configurations).

\begin{table*}[!t]
\centering
\small{
\begin{tabular}{ccll}
  \multicolumn{4}{c}{\textbf{Table~8}}\\
  \multicolumn{4}{c}{\textsc{Comparison with Previous Works}}\\
\hline\hline
 $M$ & $M/L_\mathrm{V}$& dynamical model & reference   \\
($10^5$ $M_\odot$)&($M_\odot/L_\odot$)& &  \\
\hline
$10.7\pm0.98$&$1.17^{+0.53}_{-0.43}$&spherical \citet{1975AJ.....80..175W}&\citet{2005ApJS..161..304M} \\
$9.0$        & $1.52$           &spherical Monte Carlo model&\citet{2011MNRAS.410.2698G}\\ 
$7.18\pm0.41$& $1.34\pm0.08$    &spherical \citet{1966AJ.....71...64K}&\citet{2012AA...539A..65Z}\\
$6.23\pm0.04$& $1.69\pm0.13$    &rotating \citet{VarriBertin2012}&\citet{paolo13} \\
$5.57^{+0.33}_{-0.28}$ & $1.40\pm0.03$ &isotropic Jeans model&\citet{2015ApJ...803...29W}\\
$7.00\pm0.06$ & $1.99\pm0.20$&$N$-body model&\citet{2017MNRAS.464.2174B}\\
$7.77$              & $1.63$&anisotropic $f_{T}^{(\nu)}$ model   &\citet{2016AA...590A..16D}\\
$8.4\pm0.4$ & $1.98\pm0.26$&rotating \citet{VarriBertin2012}&this work\\
\hline\hline
\end{tabular}}
\tablecomments{Compilation of the total mass and the mass-to-light
  ratio of 47~Tuc derived from dynamical modeling in previous works.}
\end{table*}

A complementary path would consist in employing empirical
  models that are optimally designed to describe the data (e.g., Jeans
  models or orbit-based Schwarzschild models). Despite providing more
  freedom in the description of the data, the major drawback is they
  do offer a very limited connection to the underlying physical
  picture of the stellar system under consideration. However, for
  studies aimed at the understanding of specific physical ingredients
  (e.g., the possible presence of a intermediate-mass black hole in
  47~Tuc, the exact mass-to-light ratio of the stellar population, or
  the amount of energy equipartition), we would certainly benefit from
  complementary and alternative modeling approaches from the one
  employed here. In addition, thanks to recent progress on the
  computational side, it has recently been proved that realistic
  N-body models, following the entire dynamical evolution of the
  system, are finally within reach, at least for selected Galactic
  globular clusters (e.g., see the model of M4 presented by
  \citealt{2014MNRAS.445.3435H} or the DRAGON series by
  \citealt{2015MNRAS.450.4070W}), although it should be kept in mind
  that the performance cost of the exploration of a wide range of
  initial conditions is still far from negligible, especially for
  massive clusters such as 47~Tuc.

In order to understand the limitations intrinsic to our
  modeling, we report in Table~8 a compilation of the total mass and
  the mass-to-light ratio for 47 Tuc~from the literature obtained
  using a variety of dynamical modeling techniques. These
  include:\ i) distribution-function-based models (spherical and
  isotropic models, \citealt{2005ApJS..161..304M,
    2012AA...539A..65Z}, spherical and anisotropic models,
  \citealt{2016AA...590A..16D}, and axisymmetric rotating models,
  \citealt{paolo13}), ii) isotropic spherical Jeans models
  (\citealt{2015ApJ...803...29W}), and iii) $N$-body and spherical
  Monte Carlo models (\citealt{2017MNRAS.464.2174B,
    2011MNRAS.410.2698G}). Note that all these previous works, with
  the exception of \citet{paolo13}, do not take into consideration
  internal rotation, therefore they are intrinsically unable to
  reproduce our state-of-the-art three-dimensional kinematic
  measurements.

From Table~8, it is evident that the estimates in the
  literature depend on the particular modeling techniques
  employed. The values of the total mass of 47~Tuc ranges between
  5.5--10.7$\times$10$^5$ $M_\odot$ with our model giving an
  intermediate value of 8.4$\times$10$^5$ $M_\odot$, while the
  mass-to-light ratio ranges from 1.2--2 $M_\odot/L_\odot$, with our
  model giving 1.98 $M_\odot/L_\odot$, consistent with the upper
  limit. Given the agreement with these previous works, we are
  confident that our model is able to reproduce the global properties
  of 47~Tuc, despite some of the discrepancies reported above.

\section{Conclusions}
\label{sec:concl}

We have derived for the first time the plane-of-the-sky rotation of
the GC 47~Tuc from the core out to 13$^\prime$ (or 3--4
  half-light radii, or $\sim$30\% of the tidal radius), together with
detailed radial and tangential velocity-dispersion profiles. The
cluster's plane-of-the-sky kinematic information is coupled to
literature LOS measurements of the same quantities and
surface-brightness profiles, and simultaneously fit with
state-of-the-art dynamical models.  From the application of the
rotating dynamical model we conclude that:
\begin{itemize}
\item{PMs are \textit{critically necessary} if we hope to constrain
  the intrinsic structure of GCs. With the application of our
  dynamical model to both kinematics and photometry we have obtained a
  full three-dimensional description of 47~Tuc;}
\item{The higher rotation in the plane of sky with respect to the one
  along the LOS implies an inclination angle of the rotation
  axis of $\simeq30^\circ$;}
\item{Our best-fit dynamical model predicts an intrinsic V/$\sigma$
  profile that reaches values of $\simeq0.9$ around two half-light
  radii from the cluster's center;}
\item{On the basis of our global dynamical analysis, we confirm that
  the observed flattening of the cluster is most likely due to its
  appreciable internal rotation, as the projected ellipticity profile
  determined by our model is in good agreement with the ellipticity
  measurements currently available for 47~Tuc. The three-dimensional
  morphological structure implied by our axisymmetric model is complex,
  and may be characterized by a non-monotonic intrinsic ellipticity
  profile, which reaches values of about 0.45 towards the intermediate
  regions of the cluster.}
\end{itemize}

This comprehensive dynamical investigation, based on new PM
measurements of unprecedented accuracy, has allowed us to unveil a new
degree of kinematic complexity in 47~Tuc. Such a superb
characterization of the three-dimensional velocity space offers a
novel and invaluable ground for the study of numerous aspect of
collisional gravitational dynamics, from the long-forgotten role
played by angular momentum to the tantalizing opportunity of finally
exploring the phase space properties of its stellar populations, with
the goal of providing a key contribution towards a more realistic
dynamical paradigm for this class of stellar systems

\acknowledgments AB acknowledges support from STScI grant AR-12845
(PI:\ Bellini). PB acknowledges financial support from CITA National
Fellowship. ALV acknowledges support from the EU Horizon 2020 program
in the form of a Marie Sklodowska-Curie Fellowship (MSCA-IF-EF-RI
658088). GP acknowledges partial support by PRIN-INAF 2014 and by the
``Progetto di Ateneo 2014 CPDA141214'' by Universit\`a di Padova.

\end{document}